\documentclass[aps,prx,reprint,superscriptaddress,floatfix]{revtex4-2}

\usepackage{amsmath}
\usepackage{amssymb}
\usepackage{bm}
\usepackage{graphicx}
\usepackage{placeins}
\usepackage{booktabs}
\usepackage{multirow}
\usepackage{xcolor}
\definecolor{currentrevision}{RGB}{0,0,0}
\colorlet{red}{black}
\usepackage{tikz}
\usepackage{etoolbox}
\usepackage[colorlinks=true,linkcolor=blue,citecolor=blue,urlcolor=blue,filecolor=blue]{hyperref}
\AtBeginDvi{}
\DeclareRobustCommand{\rev}[1]{{#1}}
\DeclareRobustCommand{\auditrev}[1]{{#1}}
\DeclareRobustCommand{\tcnew}[1]{{#1}}


\newcommand{\cfg}[1]{\ensuremath{\{#1\}}}

\newcommand{\WCC}{Wannier charge center}

\makeatletter
\begin{document}

\title{Irreducible Weyl Semimetals}

\author{Ke-Xin Pang}
\affiliation{State Key Laboratory of Metastable Materials Science and Technology and Hebei Key Laboratory of Microstructural Material Physics, School of Science, Yanshan University, Qinhuangdao 066004, China}

\author{Yan Gao}
\email{yangao9419@ysu.edu.cn}
\affiliation{State Key Laboratory of Metastable Materials Science and Technology and Hebei Key Laboratory of Microstructural Material Physics, School of Science, Yanshan University, Qinhuangdao 066004, China}

\date{\today}

\begin{abstract}
A key objective in Weyl-semimetal (WSM) research is to identify phases built from minimal configurations of Weyl points (WPs), which provide the simplest settings for investigating the intrinsic physics of chiral topological charges. Pang \textit{et al.} [\rev{Phys. Rev. Res. \textbf{8}, 033304 (2026)}] recently established that crystallographically realizable WSMs can be decomposed into linear combinations of 16 minimal “irreducible Weyl molecules” (IWMs). A fundamental question remains unresolved: which of these non-decomposable configurations can themselves form standalone crystalline phases? Here we introduce irreducible Weyl semimetals (IWSMs), in which two bands crossing near the Fermi level form an elementary symmetry-compatible complex whose complete Weyl-node configuration is charge neutral and crystallographically indivisible. Combining all 1651 magnetic space groups with Weyl-orbit multiplicities and band-compatibility relations, we classify IWSMs formed by crystalline-symmetry-protected twofold WPs. Only ten of the sixteen primitive inventories admit one-copy realizations, comprising four Pair, four Split, and two Mixed charge-node architectures. Remarkably, this classification uncovers five unconventional and previously unrecognized topological phases: the charge-three Pair IWSM $\{3, -3\}$; the Split IWSMs $\{3, -1, -1, -1\}$, and $\{4, -1, -1, -1, -1\}$; and the Mixed IWSMs $\{3, 1, -2, -2\}$ and $\{3, 3, -2, -2, -2\}$. For all ten classes, symmetry-constrained lattice models verify the complete node inventories, Chern charges, and surface chiral-flow incidence. Our results establish crystalline irreducibility as a phase-level organizing principle beyond individual Weyl nodes and provide a symmetry-resolved framework for identifying minimal Weyl complexes in electronic, phononic, and photonic systems.

\end{abstract}

\maketitle

\section{Introduction}
\label{sec:introduction}
Weyl semimetals (WSMs) are gapless topological phases in which Weyl points (WPs) act as quantized monopoles of Berry curvature in momentum space~\rev{\cite{armitage2018,lv2021rmp}}. Each WP carries an integer chiral charge $C$, while the Nielsen--Ninomiya theorem requires the total chiral charge over the Brillouin zone (BZ) to vanish~\rev{\cite{nielsen1981a,nielsen1981b}}. Consequently, the physically relevant object is not an isolated WP but a globally charge-neutral Weyl-node configuration. A central objective has therefore been to identify WSM phases built from the minimal possible Weyl configurations, for which the contributions of individual chiral nodes and their associated boundary states can be isolated most cleanly. The pursuit of a single pair of oppositely charged WPs, often viewed as the topological analogue of the hydrogen atom, is the simplest example of this program~\rev{\cite{wang2019eucd2as2,zheng2026boron,jin2017ferromagnetic,Gao2023MnX2B2T6,taddei2022canted,bai2026sp2carbon,ding2023baniio6,qian2026v3s4}}. At the same time, crystalline symmetries can stabilize multi-Weyl points carrying $|C|>1$~\rev{\cite{fang2012multiweyl}}, substantially enlarging the possible minimal charge-neutral configurations beyond the conventional $\{1,-1\}$ pair~\cite{pang2026universal}.

Minimality at the level of chiral charge, however, does not by itself imply minimality of a crystalline Weyl phase. Crystal and magnetic symmetries act on a WP by generating complete momentum-space orbits, while nonsymmorphic and antiunitary symmetries can impose additional band connectivity and degeneracies~\rev{\cite{watanabe2018msg,elcoro2021mtqc,bradlyn2017}}. A primitive neutral charge inventory may therefore require additional symmetry-related Weyl partners or additional closures of the same crossing-band gap and thus cannot necessarily exist as an independent crystalline phase. Existing symmetry classifications largely address the local question of which topological quasiparticles and chiral charges are allowed by a given space group or magnetic space group~\cite{Yu2022TypeII,Liu2022TypeIII,Zhang2022TypeIV}. They do not, by themselves, determine whether a prescribed minimal charge inventory can constitute the complete Weyl content of an independent crystalline phase.

Recently, Pang \rev{\textit{et al.}} proposed that every crystallographically realizable twofold-Weyl configuration within the stable charge domain $|C|\leq4$ can be decomposed into non-negative-integer combinations of 16 irreducible Weyl molecules (IWMs)~\rev{\cite{pang2026universal}}. These IWMs constitute the minimal charge-neutral building blocks of Weyl matter and are primitive with respect to chiral-charge decomposition. Charge-space primitivity, however, does not guarantee that an IWM can itself form a complete, standalone crystalline Weyl phase. A fundamental question therefore remains unanswered: which of these primitive charge-neutral modules can themselves exist as standalone, non-decomposable Weyl phases in real crystals? Equivalently, when does charge-level irreducibility survive the complete action of crystalline symmetry and band connectivity, so that one primitive Weyl inventory exhausts the crossings formed by the relevant pair of bands near the Fermi level?

Here we answer this question by introducing the concept of an irreducible Weyl semimetal (IWSM). An IWSM is a Weyl phase in which two crossing bands near the Fermi level form an elementary symmetry-compatible band complex whose complete Weyl-node configuration is both charge neutral and crystallographically indivisible. This requires three distinct conditions: the Weyl-charge inventory must be primitive; complete magnetic-space-group (MSG) orbits must reproduce exactly one copy of that inventory without compulsory additional Weyl partners; and all constituent crossings must belong consistently to the same symmetry-compatible band complex without band connectivity forcing additional crossings. The distinction is essential: an IWM defines an elementary module of the universal charge algebra, whereas an IWSM is a complete crystalline phase realization of such a module.

We apply these conditions to crystalline-symmetry-protected WPs across all 1651 MSGs, treating the regimes without and with spin-orbit coupling (SOC) separately and combining exact Weyl-orbit multiplicities with global band connectivity and irreducible-representation/corepresentation compatibility. Of the 16 primitive IWM inventories, exactly ten admit one-copy standalone realizations. They organize into four Pair, four Split, and two Mixed charge-node architectures. The classification reveals \rev{five} unconventional irreducible Weyl-phase classes not previously identified as standalone phases: the Pair IWSM $\{3,-3\}$; the Split IWSMs $\{3,-1,-1,-1\}$ \rev{and} $\{4,-1,-1,-1,-1\}$; and the Mixed IWSMs $\{3,1,-2,-2\}$ and $\{3,3,-2,-2,-2\}$. A notable global consequence is that none of the ten IWSM classes survives in spinful Type-II MSGs with SOC. For every allowed IWSM class, we further construct an explicit symmetry-constrained lattice Hamiltonian and verify its complete BZ Weyl-node inventory, symmetry representations, and quantized chiral charges. The associated surface spectra demonstrate the net chiral-flow incidence required by the projected bulk charges and expose three distinct boundary architectures: one-to-one Pair, one-to-many or many-to-one Split, and many-to-many Mixed networks. These results elevate Weyl minimality from a local node-counting problem to a phase-level factorization problem and establish crystalline irreducibility as a general organizing principle for minimal Weyl complexes. Although formulated here for electronic WSMs near the Fermi level, the underlying charge, symmetry, and compatibility structure provides a natural route for identifying analogous irreducible chiral complexes in phononic and photonic band systems.

\section{The irreducible Weyl phase}
\label{sec:irreducible_phase}

\subsection{Definition and operational criterion}
\label{sec:definition_criterion}

\begin{figure}[!tbp]
	\centering
	\includegraphics[width=\columnwidth]{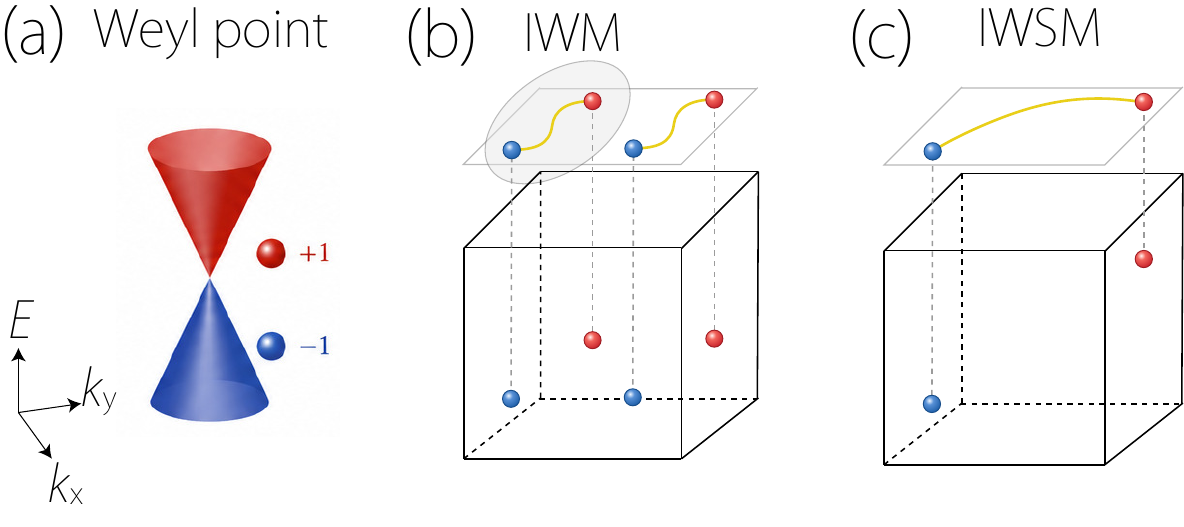}
	\caption{
		Schematic illustration of Weyl points, irreducible Weyl molecules, and irreducible Weyl semimetals.
		(a) A Weyl point (WP) is a local twofold crossing carrying quantized chiral charge.
		(b) Irreducible Weyl molecules (IWMs) are the elementary charge-neutral building blocks of Weyl semimetals. The shaded ellipse highlights one IWM, whose node configuration contains no nonempty proper charge-neutral subset.
		(c) An irreducible Weyl semimetal (IWSM) realizes one IWM as \rev{the complete zero set of a selected adjacent-band gap throughout the BZ}.
		\auditrev{In (a), the $+1$ and $-1$ markers denote the opposite Chern numbers of the two crossing bands; in (b) and (c), red and blue spheres denote WPs of opposite chirality.}
		Dashed lines indicate surface projections, and yellow curves schematically represent admissible Fermi-arc connections.
	}
	\label{fig:definition}
\end{figure}

The distinction between an irreducible Weyl molecule (IWM) and an irreducible Weyl semimetal (IWSM) requires three levels of description to be kept separate. A Weyl point (WP) is a local band crossing carrying a quantized chiral charge $C$ [\rev{Fig.~\ref{fig:definition}(a)}]. At the level of charge algebra, an IWM is a primitive charge-neutral Weyl-node inventory~\cite{pang2026universal}: its total chiral charge vanishes, whereas no proper subset is itself charge neutral [\rev{Fig.~\ref{fig:definition}(b)}]. The irreducibility of an IWM is therefore algebraic and does not by itself imply that the corresponding charge inventory can constitute an independent crystalline Weyl phase.

Crystalline irreducibility must instead be defined for a complete pair of crossing bands. For a selected pair of adjacent bands, we introduce the momentum-dependent band separation
\begin{equation}
	g_N(\mathbf{k})
	=
	E_{N+1}(\mathbf{k})-E_N(\mathbf{k}),
	\label{eq:band_separation}
\end{equation}
where $E_N(\mathbf{k})$ and $E_{N+1}(\mathbf{k})$ are the energies of the two active bands and $N$ is their band index. In an electronic realization, this active band pair is the pair of bands forming the Weyl phase near the Fermi level. A Weyl crossing of this pair occurs at a momentum $\mathbf{k}_j$ for which
$g_N(\mathbf{k}_j)=0$.
The constituent WPs need not be exactly isoenergetic; what defines the same Weyl sector is that they arise from the same pair of adjacent bands.

We denote by
\begin{equation}
	\mathcal{Z}_N
	=
	\left\{
	\mathbf{k}\in{\rm BZ}
	\,\middle|\,
	g_N(\mathbf{k})=0
	\right\}
	\label{eq:zero_set}
\end{equation}
the complete set of momenta at which this selected band pair becomes degenerate. For an IWSM, every element of $\mathcal{Z}_N$ must be an isolated WP belonging to one primitive IWM inventory. Explicitly, if
$\mathcal{Z}_N=\{\mathbf{k}_j\}_{j=1}^{N_{\rm WP}}$
and $C_j$ is the chiral charge of the WP at $\mathbf{k}_j$, then
\begin{equation}
	\mathcal{Z}_N
	=
	\{\mathbf{k}_j\}_{j=1}^{N_{\rm WP}},
	\qquad
	\mathcal{C}_N
	\equiv
	\{C_j\}_{j=1}^{N_{\rm WP}}
	=
	X_{\rm IWM},
	\label{eq:iwsm_definition}
\end{equation}
where $X_{\rm IWM}$ denotes exactly one primitive IWM charge inventory, including all symmetry-required members of its complete magnetic-space-group orbit. We refer to Eq.~(\ref{eq:iwsm_definition}) as a \emph{one-copy IWM realization}. Thus, within the selected adjacent-band pair, the complete set of gapless crossings is precisely one IWM rather than an integer multiple or a superposition of several neutral Weyl complexes [\rev{Fig.~\ref{fig:definition}(c)}].

Equation~(\ref{eq:iwsm_definition}) defines irreducibility for the active two-band Weyl sector. Degeneracies involving other adjacent-band pairs, $g_m(\mathbf{k})$ with $m\neq N$, do not alter this classification because they do not belong to the selected crossing-band complex. If additional Weyl crossings from other band pairs also occur near the Fermi level, the selected pair can still realize an irreducible Weyl sector, whereas the full material is more appropriately regarded as an IWSM-containing Weyl metal rather than an isolated IWSM. In the strict standalone-IWSM limit considered below, the low-energy Weyl physics is dominated by the selected irreducible band pair. This irreducibility is preserved under symmetry-compatible deformations that keep $g_N(\mathbf{k})\neq0$ away from the constituent WPs while preserving their chiral charges, complete MSG orbits, and protecting symmetries. Changing the factorization therefore requires a change in the zero set $\mathcal{Z}_N$, in the signed Weyl-node inventory, or in the protecting crystalline symmetry.

Operationally, an IWSM must satisfy three conditions. First, its signed Weyl-charge inventory must be primitive, excluding any proper charge-neutral subinventory. Second, complete magnetic-space-group (MSG) orbits must reproduce exactly one copy of that inventory without compulsory additional Weyl partners. Third, all constituent crossings must belong consistently to one symmetry-compatible band complex, with global band connectivity and irreducible-representation or magnetic-corepresentation compatibility introducing no additional zeros of $g_N(\mathbf{k})$. These conditions test, respectively, charge primitivity, crystalline orbit closure, and band realizability.

\subsection{Crystalline symmetry criteria and classification}
\label{sec:classification}

We now determine which of the 16 primitive IWM inventories satisfy these stronger crystalline conditions. Within the stable twofold-Weyl domain, the allowed local chiral charges are
\begin{equation}
	\mathcal{G}_0
	=
	\{\pm1,\pm2,\pm3,\pm4\},
	\label{eq:charge_alphabet}
\end{equation}
where $|C|=4$ is the maximal crystallographically stable chiral charge of a twofold WP in three-dimensional crystals~\cite{Yu2022TypeII,Liu2022TypeIII,Zhang2022TypeIV,zhang2020quadruple,cui2021chargefour}. For a primitive charge inventory
\begin{equation}
	X
	=
	\{C_1,C_2,\ldots,C_{N_{\rm WP}}\},
\end{equation}
we introduce the eight-component signed charge-count vector
\begin{equation}
	\boldsymbol{\nu}_X
	=
	\left(
	N_{+4},
	N_{+3},
	N_{+2},
	N_{+1},
	N_{-1},
	N_{-2},
	N_{-3},
	N_{-4}
	\right)^{\mathsf T},
	\label{eq:inventory_vector}
\end{equation}
where $N_C$ denotes the number of WPs carrying chiral charge $C$. Defining
\begin{equation}
	\mathbf{c}
	=
	\left(
	4,3,2,1,-1,-2,-3,-4
	\right)^{\mathsf T},
\end{equation}
the Nielsen--Ninomiya charge-neutrality condition becomes
\begin{equation}
	\mathbf{c}^{\mathsf T}\boldsymbol{\nu}_X
	=
	\sum_C C N_C
	=
	\sum_{j=1}^{N_{\rm WP}} C_j
	=
	0.
	\label{eq:charge_neutrality}
\end{equation}
The inventory $X$ is primitive when no nonzero proper subinventory is itself charge neutral, i.e.,
\begin{equation}
	\begin{gathered}
	\nexists\,\boldsymbol{\mu}\in\mathbb{Z}_{\geq0}^{8}\quad {\rm such~that}\\
	\rev{\mathbf{0}\leq\boldsymbol{\mu}\leq\boldsymbol{\nu}_X,\quad
	\boldsymbol{\mu}\notin\{\mathbf{0},\boldsymbol{\nu}_X\}},\\
	\mathbf{c}^{\mathsf T}\boldsymbol{\mu}=0,
	\end{gathered}
	\label{eq:primitivity}
\end{equation}
where the inequalities are understood componentwise \rev{and the excluded endpoints ensure that the subinventory is nonzero and proper}. Equations~(\ref{eq:charge_neutrality}) and (\ref{eq:primitivity}) define the primitive charge-neutral IWM inputs~\rev{\cite{pang2026universal}}. They do not yet guarantee a standalone crystalline realization. The local symmetry support is documented in the Supplemental Material (SM)~\cite{supplemental}, Sec.~I\,B\,1 and Table~S2. Complete crystalline realizability requires the two further tests below.

For an MSG operation $g=\{R_g|\boldsymbol\tau_g\}\mathcal T^{\epsilon_g}$, the momentum and Weyl charge transform as
\begin{equation}
 \begin{gathered}
 g\mathbf k=(-1)^{\epsilon_g}R_g\mathbf k\pmod{\mathcal L^*},\\
 C(g\mathbf k)=\chi(g)C(\mathbf k),\qquad \chi(g)=\det R_g .
 \end{gathered}
 \label{eq:msg_charge_action}
\end{equation}
Here $R_g$ acts in Cartesian coordinates, $\epsilon_g=0,1$ distinguishes unitary and antiunitary operations, and $\mathcal L^*$ is the reciprocal lattice. In particular, time reversal preserves chirality, whereas an improper spatial operation reverses it, whether or not it is combined with time reversal. Fractional translations and spin-dependent phases enter the representations and compatibility relations, but not this charge-transformation rule~\cite{Yu2022TypeII,Liu2022TypeIII,Zhang2022TypeIV,Alpin2023}.

Let $\overline G=G/L$ be the finite group obtained by removing lattice translations, and let $\overline G_{\mathbf k}$ stabilize $\mathbf k$ modulo $\mathcal L^*$. The complete momentum star has cardinality
\begin{equation}
 m(\mathbf k)=[\overline G:\overline G_{\mathbf k}]
 =\frac{|\overline G|}{|\overline G_{\mathbf k}|}.
 \label{eq:star_cardinality}
\end{equation}
Every charged WP must satisfy $\overline G_{\mathbf k}\subseteq\overline G^+$, where $\overline G^+=\ker\chi$. An improper stabilizer would require $C=-C$ and thus exclude a nonzero Weyl charge. If $\overline G$ contains any chirality-reversing operation, the entire inventory must obey $N_c=N_{-c}$ for each magnitude $c$. The subgroup $\overline G^+$ can contain antiunitary operations; it is not the unitary subgroup of an MSG.

The local starting points are the symmetry-protected twofold WP channels at high-symmetry points (HSPs) and high-symmetry lines (HSLs) identified in the emergent-particle catalogues~\cite{Tang2021EffectiveModels,Yu2022TypeII,Liu2022TypeIII,Zhang2022TypeIV}. We restrict the classification to these crystalline-symmetry-protected channels, including for $|C|=1$. A topologically stable unit-charge WP at a general momentum need not be protected or pinned by crystalline symmetry and lies outside this screening domain. For every admitted channel, the local irrep or magnetic corepresentation must allow an isolated twofold crossing with the specified charge in the chosen spin convention. Multifold degeneracies and nodal lines or planes do not supply such a channel.
 
The first crystalline test is exact MSG-orbit closure. For a given MSG $G$ and spin convention $\zeta$, we denote by
$\Omega_{G,\zeta}$
the set of symmetry-inequivalent Weyl-orbit types allowed by the local little-group symmetry. Here
$\zeta={\rm SV}$
denotes the single-valued representation convention appropriate to spinless systems or an effectively decoupled spin channel, whereas
$\zeta={\rm DV}$
denotes double-valued representations appropriate to spinful electronic systems with SOC.

Each orbit type
$\alpha\in\Omega_{G,\zeta}$
is specified by a representative Weyl point
$(\mathbf{k}_{\alpha},C_{\alpha})$
together with all Weyl points generated from it by the unitary and antiunitary operations of $G$. We define
\begin{equation}
	\begin{split}
	\mathbf{m}_{G,\zeta,\alpha}={}&\big(
	M^{(\alpha)}_{+4},M^{(\alpha)}_{+3},M^{(\alpha)}_{+2},M^{(\alpha)}_{+1},\\
	& M^{(\alpha)}_{-1},M^{(\alpha)}_{-2},M^{(\alpha)}_{-3},M^{(\alpha)}_{-4}
	\big)^{\mathsf T}
	\end{split}
	\label{eq:orbit_inventory}
\end{equation}
as the signed charge-count vector of this complete Weyl orbit, where
$M^{(\alpha)}_C$
is the number of symmetry-related WPs with charge $C$ generated by the orbit type $\alpha$.

A primitive IWM inventory $X$ can form a one-copy crystalline realization only if there exists at least one pair $(G,\zeta)$ and a set of non-negative integers $n_\alpha$ satisfying
\begin{equation}
	\sum_{\alpha\in\Omega_{G,\zeta}}
	n_{\alpha}\,
	\mathbf{m}_{G,\zeta,\alpha}
	\in
	\left\{
	\boldsymbol{\nu}_X,\,
	\mathcal{R}\boldsymbol{\nu}_X
	\right\},
	\qquad
	n_{\alpha}\in\mathbb{Z}_{\geq0}.
	\label{eq:orbit_closure}
\end{equation}
Here $n_\alpha$ is the number of complete Weyl orbits of type $\alpha$ used to construct the phase, and $\mathcal{R}$ denotes global chirality reversal,
\begin{equation}
	\begin{split}
		\mathcal{R}\big(&N_{+4},N_{+3},N_{+2},N_{+1},\\
		&N_{-1},N_{-2},N_{-3},N_{-4}\big)^{\mathsf T}\\
		={}&\big(N_{-4},N_{-3},N_{-2},N_{-1},\\
		&N_{+1},N_{+2},N_{+3},N_{+4}\big)^{\mathsf T}.
	\end{split}
	\label{eq:chirality_reversal}
\end{equation}
Equation~(\ref{eq:orbit_closure}) is stronger than charge neutrality and stronger than requiring that an MSG can host the necessary charge magnitudes separately. It requires all signed multiplicities of the IWM to be supplied simultaneously and exactly by complete MSG orbits. If Eq.~(\ref{eq:orbit_closure}) has no solution for every MSG $G$ and both spin conventions $\zeta$, the corresponding IWM is excluded as a one-copy IWSM independently of any Hamiltonian parameters. \rev{SM~\cite{supplemental} Sec.~I\,B\,2 and Table~S3 record the exact-orbit support, before target-gap compatibility is imposed.}

The unit-charge Pair inventory $\{1,-1\}$ requires both a global two-node test and local crystalline protection. Chern-slice integration relates the Chern vector of the lower-$N$ projector to the separation of the opposite-charge nodes modulo a reciprocal-lattice vector~\cite{yang2011}. A distinct pair therefore requires a nonzero symmetry-allowed time-odd axial vector. \tcnew{Within this set, we retain both distinct-branch rotation/screw HSL crossings and symmetry-pinned HSP Weyl degeneracies that close in the same adjacent gap. The combined HSP/HSL classification gives 45 Type-I and 193 Type-III Belov--Neronova--Smirnova (BNS) settings, or 238 settings in each spin convention [SM~\cite{supplemental}, Sec.~I\,A\,2 and Tables~S4 and S16].} General-momentum pairs and same-representation crossings that can leave the designated HSL are outside the present catalogue. The higher-charge inventories are screened using the same symmetry-protected HSP/HSL domain.

Exact MSG-orbit closure is necessary but not sufficient. The second crystalline test is whether all symmetry-allowed Weyl crossings can belong to one common pair of adjacent bands without crystalline band connectivity forcing additional crossings of that same pair. This condition is determined by the compatibility relations between the irreducible representations or magnetic corepresentations of the active bands.

Consider a high-symmetry line $\ell$ connecting two high-symmetry momenta $\mathbf{k}_1$ and $\mathbf{k}_2$. Let
$G_{\mathbf{k}_1}$ and $G_{\mathbf{k}_2}$
be the corresponding endpoint little groups and
$G_{\ell}$
the little group of the line, with
$G_{\ell}\subseteq G_{\mathbf{k}_i}$.
Let
$\rho^{\rm act}_{\mathbf{k}_i}$
denote the irrep or magnetic corepresentation carried by an active band at the endpoint $\mathbf{k}_i$. Upon restriction (subduction) to the line little group,
\begin{equation}
	\operatorname{Res}^{G_{\mathbf{k}_i}}_{G_{\ell}}
	\rho^{\rm act}_{\mathbf{k}_i}
	=
	\bigoplus_{\beta}
	a_{i\beta}\,
	\rho_{\ell,\beta},
	\qquad
	i=1,2,
	\label{eq:compatibility}
\end{equation}
where
$\rho_{\ell,\beta}$
denotes the $\beta$th irrep or magnetic corepresentation of $G_\ell$, and
$a_{i\beta}\in\mathbb{Z}_{\geq0}$
is its branching multiplicity in the restriction of
$\rho^{\rm act}_{\mathbf{k}_i}$.
Two endpoint branches can connect continuously along $\ell$ only through line representations that occur in the corresponding restrictions~\rev{\cite{bradlyn2017,watanabe2018msg,elcoro2021mtqc}}. The complete IWSM screening therefore follows all active HSP/HSL branches and their MSG stars and retains a setting only when the required crossings can be embedded consistently in the same selected adjacent-band pair without an incompatible branch termination or an unavoidable additional zero of $g_N(\mathbf{k})$. The final target-gap check also excludes additional accidental crossings away from these symmetry channels.

\begin{figure*}[t]
\centering
\includegraphics[width=\textwidth]{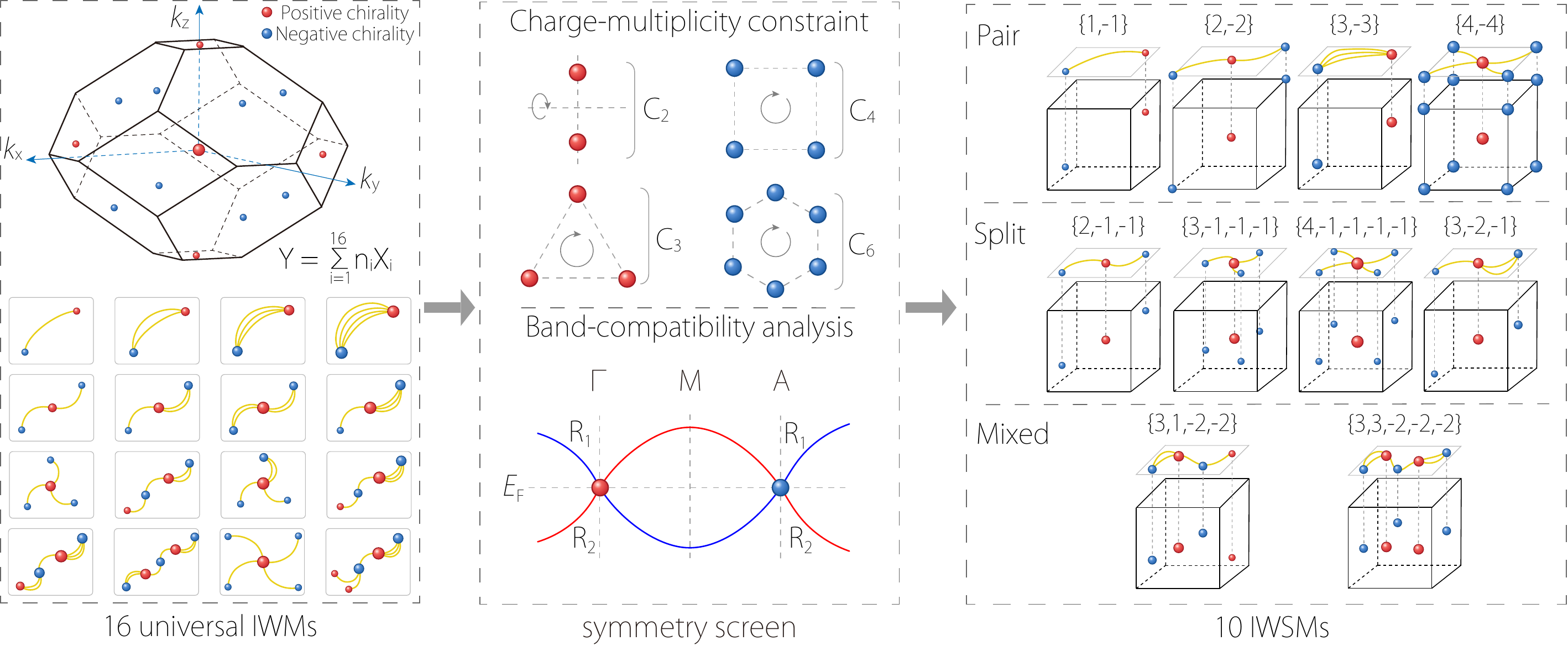}
\caption{
	Crystalline screening of the sixteen IWM configurations~\rev{\cite{pang2026universal}}.
	Exact magnetic-space-group (MSG) orbit closure and band-representation compatibility determine whether an IWM can exhaust the Weyl crossings of one selected adjacent-band pair.
	Ten configurations admit standalone IWSM realizations: four Pair, four Split, and two Mixed classes.
	For the remaining six configurations, crystalline symmetry requires additional Weyl nodes.
	Red and blue spheres denote opposite chiralities, yellow curves illustrate surface connections, and $E_{\rm F}$ marks the Fermi level in the schematic band-compatibility analysis.
}
\label{fig:reduction}
\end{figure*}

The two crystalline tests address distinct physical questions [\rev{Fig.~\ref{fig:reduction}}]. Equation~(\ref{eq:orbit_closure}) asks whether crystalline symmetry can supply exactly the signed Weyl-node multiplicities of one IWM. Equation~(\ref{eq:compatibility}) asks whether these symmetry-allowed nodes can belong to one common crossing-band complex. A primitive IWM can therefore fail to become an IWSM either because MSG symmetry enforces an incompatible orbit multiplicity or because the allowed representations enforce additional crossings of the same active band pair. The complete condition also requires the absence of accidental extra zeros throughout the BZ, which is verified for the explicit lattice realizations.

The orbit criterion has a particularly transparent form for the four Pair classes $\{q,-q\}$, with $q=1,2,3,4$. There are exactly two possibilities. If every spatial operation is proper, each WP must form a singleton star, and the two opposite-charge nodes occupy symmetry-inequivalent positions. If a chirality-reversing operation is present, the two WPs form one two-arm star. Its stabilizer equals $\overline G^+$ and has index two in $\overline G$; an operation in the other coset exchanges the opposite charges. Both mechanisms require locally allowed charge-$q$ twofold channels and a common compatible target gap. For $q=4$, a multifold crossing with a band Chern number of four is not a substitute for a twofold charge-four WP.

The four Split and two Mixed inventories obey a stronger common condition. None has $N_c=N_{-c}$ for every $c$, so all spatial operations of its MSG must be proper. Consequently, every star is monochiral. Let $m_c$ denote a complete $m$-arm star whose nodes each carry charge $c$, and let $\oplus$ join distinct stars. The allowed star partitions are then
\begin{align}
 \{2,-1,-1\}:&\quad 1_{+2}\oplus(2_{-1}\ \text{or}\ 1_{-1}\oplus1_{-1}),\nonumber\\
 \{3,-1,-1,-1\}:&\quad 1_{+3}\oplus\mathcal P_3(-1),\nonumber\\
 \{4,-1,-1,-1,-1\}:&\quad 1_{+4}\oplus\mathcal P_4(-1),\nonumber\\
 \{3,-2,-1\}:&\quad 1_{+3}\oplus1_{-2}\oplus1_{-1},\nonumber\\
 \{3,1,-2,-2\}:&\quad 1_{+3}\oplus1_{+1}\oplus\mathcal P_2(-2),\nonumber\\
 \{3,3,-2,-2,-2\}:&\quad\mathcal P_2(+3)\oplus\mathcal P_3(-2).
 \label{eq:ten_star_partitions}
\end{align}
Here $\mathcal P_n(c)$ ranges over the integer partitions of $n$ into complete same-charge stars. Explicitly, $\mathcal P_2$ permits $2$ or $1+1$ arms; $\mathcal P_3$ permits $3$, $2+1$, or $1+1+1$; and $\mathcal P_4$ permits $4$, $3+1$, $2+2$, $2+1+1$, or $1+1+1+1$. Thus the unit-charge terminals of a Split IWSM need not all be symmetry equivalent. Likewise, the two charge-three nodes of $\{3,3,-2,-2,-2\}$ need not be singleton stars. Its two-arm/three-arm choice requires little-group indices two and three within the same MSG, so $|\overline G|$ must be divisible by six.

These partitions specify the required orbit structure, not independent choices of local channels. Every star must have the actual index in Eq.~(\ref{eq:star_cardinality}), and all its arms must be counted once. Different stars must coexist in one MSG and satisfy the same target-gap compatibility conditions. The prescribed node configuration must persist under sufficiently small symmetry-preserving perturbations. A momentum star is in one-to-one correspondence with the left cosets of its stabilizer. The stabilizer need not be normal, so an index-three or index-four star does not require a quotient group of that order. These conditions, applied to the local symmetry catalogues~\cite{Yu2022TypeII,Liu2022TypeIII,Zhang2022TypeIV}, yield the setting-resolved candidates in SM~\cite{supplemental}, Tables~S4--S13. \auditrev{Global realizability additionally constrains the charge-weighted node positions through the symmetry-allowed Chern vector~\cite{yang2011}. This excludes locally allowed assignments in the higher-charge catalogues, as demonstrated in SM~\cite{supplemental}, Sec.~I\,A\,3.} The final compatible channels are collected in SM~\cite{supplemental}, Sec.~V.
 
The minimal $\{1,-1\}$ inventory illustrates the role of antiunitary symmetry. In a spinful Type-II MSG, pure time reversal $\mathcal{T}$ satisfies $\mathcal{T}^2=-1$. For a generic WP at $\mathbf{k}$, time reversal generates a distinct WP at $-\mathbf{k}$ with the same chiral charge,
\begin{equation}
	C(-\mathbf{k})=C(\mathbf{k}),
	\label{eq:TR_chirality}
\end{equation}
so charge neutrality requires a second time-reversed pair of opposite chirality~\rev{\cite{belopolski2017four,chang2018chiral}}. A generic spinful Type-II realization therefore contains at least four unit-charge WPs and cannot realize the one-copy $\{1,-1\}$ IWM. Placing opposite charges at time-reversal-invariant momenta does not restore an isolated two-node realization because Kramers degeneracy forces additional closures of the same odd-indexed adjacent-band separation. Consequently, no spinful Type-II MSG with SOC realizes a standalone $\{1,-1\}$ IWSM. The same band-pair-completeness condition eliminates the apparent Type-II-with-SOC candidates for several higher-charge inventories, including $\{3,-3\}$, $\{3,-1,-1,-1\}$, and $\{3,1,-2,-2\}$, because their charge-three Kramers--Weyl channels coexist with additional symmetry-enforced crossings of the same active band pair [see \rev{Secs.~I and V of the SM~\cite{supplemental}}].

A complementary orbit-multiplicity obstruction is provided by the primitive inventory $\{4,-2,-2\}$. Although charge-four and charge-two WPs are individually symmetry allowed, the MSGs that support the required charge-four sector do not supply the compensating charge-two sector with the exact signed multiplicity required by a single copy of $\{4,-2,-2\}$. Equation~(\ref{eq:orbit_closure}) therefore has no one-copy solution. This example demonstrates that crystalline irreducibility is a property of the complete signed Weyl inventory rather than of an isolated high-charge WP.

Applying the exact-orbit and band-compatibility conditions to all 16 primitive IWM inventories yields exactly ten IWSM classes:
\begin{align}
	\mathcal{L}_{\rm IWSM}
	=
	\{&
	\{1,-1\},
	\{2,-2\},
	\{3,-3\},
	\{4,-4\},
	\nonumber\\
	&
	\{2,-1,-1\},
	\{3,-1,-1,-1\},
	\nonumber\\
	&\{4,-1,-1,-1,-1\},
	\{3,-2,-1\},
	\nonumber\\
	&
	\{3,1,-2,-2\},
	\{3,3,-2,-2,-2\}
	\}.
	\label{eq:iwsm_classes}
\end{align}
The remaining six primitive IWM inventories,
\begin{align}
	\mathcal{L}_{\rm obs}
	=
	\{&
	\{4,-3,-1\},
	\{4,-2,-2\},
	\{4,2,-3,-3\},
	\nonumber\\
	&
	\{4,1,-3,-2\},
	\{4,-2,-1,-1\},
	\nonumber\\
	&\{4,1,1,-3,-3\}
	\},
	\label{eq:obstructed_classes}
\end{align}
fail the exact one-copy orbit condition in all 1651 MSGs in both spin conventions. They remain valid primitive charge-neutral modules of the IWM decomposition, but their crystalline realization necessarily requires additional Weyl orbits and therefore forms a larger composite WSM rather than a standalone IWSM. \rev{The six multiplicity obstructions are detailed in SM~\cite{supplemental} Sec.~I\,C and Table~S14.}

For every class in Eq.~(\ref{eq:iwsm_classes}), at least one MSG assignment satisfies both exact orbit closure and band compatibility. We further construct an explicit symmetry-constrained lattice Hamiltonian for each surviving class, perform a full-BZ scan of the selected adjacent-band pair to verify that its complete zero set contains no additional crossings, and determine the signed chiral charges from Wannier charge center winding. These constructions provide explicit existence witnesses for all ten IWSM classes within the stated twofold-Weyl domain; they do not imply that every material belonging to a compatible MSG realizes the corresponding phase. \rev{The screening evidence is consolidated in SM~\cite{supplemental} Sec.~I\,B and Table~S1. The resulting setting counts and representative materials are listed in Table~\ref{tab:iwsm-msg-reports}.}

The ten IWSMs naturally organize into four Pair inventories,
\begin{equation}
	\{1,-1\},\qquad
	\{2,-2\},\qquad
	\{3,-3\},\qquad
	\{4,-4\},
\end{equation}
four Split inventories,
\begin{equation}
	\begin{gathered}
	\{2,-1,-1\},\qquad \{3,-1,-1,-1\},\\
	\{4,-1,-1,-1,-1\},\qquad \{3,-2,-1\},
	\end{gathered}
\end{equation}
and two Mixed inventories,
\begin{equation}
	\{3,1,-2,-2\},
	\qquad
	\{3,3,-2,-2,-2\}.
\end{equation}
This organization already shows that the local monopole charge of an individual WP is insufficient to determine the crystalline Weyl phase. For example, both $\{4,-4\}$ and $\{4,-1,-1,-1,-1\}$ contain a $|C|=4$ WP, yet the Pair inventory $\{4,-4\}$ has no compatible realization with SOC, whereas the Split inventory $\{4,-1,-1,-1,-1\}$ survives in magnetic settings with SOC. Their different realizability originates from the symmetry completion of the compensating Weyl sector rather than from the local charge-four node itself.

A particularly simple global consequence emerges from the antiunitary sectors: none of the ten IWSM classes survives in a spinful Type-II MSG with SOC. This does not prohibit WPs in nonmagnetic spinful crystals. Rather, pure time reversal necessarily generates additional symmetry-related partners and/or Kramers-enforced crossings of the same active band pair, preventing a primitive IWM inventory from closing into an isolated one-copy IWSM. Crystalline irreducibility is therefore controlled by the symmetry completion and band connectivity of the entire Weyl complex, not by the local existence of an individual Weyl species.

\begin{table*}[t]
\caption{
	Magnetic-space-group (MSG) classification and representative material platforms for the ten irreducible Weyl semimetal (IWSM) classes.
	The classes are grouped into Pair, Split, and Mixed architectures.
	Entries under MSG Types I--IV count distinct BNS settings in the format ``without SOC/with SOC,'' where SOC denotes spin-orbit coupling.
	A setting is counted once if at least one assignment satisfies exact MSG-orbit closure together with global band-connectivity and irrep/corepresentation compatibility for the complete Weyl configuration of a selected adjacent-band pair, without symmetry-enforced additional crossings of that pair. Only crystalline-symmetry-protected HSP/HSL Weyl channels are included.
	Alternative node locations, Weyl-orbit assignments, or representation/corepresentation combinations within the same BNS setting are not counted separately.
	These numbers therefore represent setting-level symmetry admissibility rather than symmetry-enforced realization in every material belonging to the corresponding MSG.
		The final column lists representative reported or predicted material platforms realizing Weyl phases composed of a single irreducible Weyl molecule (IWM) configuration, as identified from their published Weyl-node inventories.
	Unmarked entries denote electronic systems; ${}^{*}$ marks candidate topological phonon materials, ${}^{\dagger}$ field-induced or spin-aligned configurations, and ${}^{\ddagger}$ electronic realizations whose relevant crossings become weakly gapped when SOC is included.
	An em dash in the material column indicates that no corresponding material example was identified among the cited reports, rather than that the IWSM class is symmetry forbidden.
	\rev{Detailed MSG and irrep/corepresentation assignments are provided in Tables~S16--S25 of the SM~\cite{supplemental}.}
}
\label{tab:iwsm-msg-reports}
\begingroup
\setlength{\tabcolsep}{3pt}
\begin{ruledtabular}
\begin{tabular}{llccccc p{0.31\textwidth}}
Motif &
IWSM inventory class &
\begin{tabular}[c]{@{}c@{}}Type I\\ settings\end{tabular} &
\begin{tabular}[c]{@{}c@{}}Type II\\ settings\end{tabular} &
\begin{tabular}[c]{@{}c@{}}Type III\\ settings\end{tabular} &
\begin{tabular}[c]{@{}c@{}}Type IV\\ settings\end{tabular} &
\begin{tabular}[c]{@{}c@{}}Total\\ settings\end{tabular} &
Representative materials \\
\hline
\multirow{4}{*}{Pair} &
\cfg{1,-1} & \tcnew{45/45} & 0/0 & \tcnew{193/193} & 0/0 & \tcnew{238/238} &
\begin{tabular}[t]{@{}l@{}}EuCd$_2$As$_2$$^{\dagger}$~\cite{soh2019ideal,wang2019eucd2as2};\\ \rev{MnSn$_2$Sb$_2$Te$_6$-B~\cite{Gao2023MnX2B2T6}}\end{tabular} \\
&
\cfg{2,-2} & \tcnew{26/17} & 31/0 & \tcnew{49/29} & \tcnew{16/6} & \tcnew{122/52} &
\begin{tabular}[t]{@{}l@{}}HDSBC-B$_{20}$$^{\ddagger}$; CR-B$_{12}$$^{\ddagger}$~\cite{zheng2026boron};\\ HSRDC-C$_{28}$~\cite{bai2026sp2carbon}\end{tabular} \\
& \cfg{3,-3} & \auditrev{8/8} & 0/0 & 14/14 & 0/0 & \auditrev{22/22} & --- \\
& \cfg{4,-4} & 4/0 & 7/0 & 0/0 & 0/0 & 11/0 & BeH$_2$$^{*}$~\cite{yang2023beh2} \\
\hline
\multirow{4}{*}{Split} &
\cfg{2,-1,-1} & 25/16 & 39/0 & 39/25 & 63/29 & 166/70 &
$\alpha$-SiO$_2$$^{*}$~\cite{wang2020triangular}; BaZnO$_2$$^{*}$~\cite{ding2022chargetwo}; ISTBN-B$_{36}$$^{\ddagger}$, CT-B$_{28}$$^{\ddagger}$, CH-B$_{36}$$^{\ddagger}$~\cite{Zheng2026TNWC}; DZQH-C$_{36}$~\cite{Bai2026Triplet} \\
& \cfg{3,-1,-1,-1} & 8/8 & 0/0 & 5/5 & \tcnew{17/17} & \tcnew{30/30} & --- \\
& \cfg{4,-1,-1,-1,-1} & 4/2 & 0/0 & 0/0 & \tcnew{5/3} & \tcnew{9/5} & --- \\
& \cfg{3,-2,-1} & 5/5 & 0/0 & 5/5 & 0/0 & 10/10 & --- \\
\hline
\multirow{2}{*}{Mixed} &
\cfg{3,+1,-2,-2} & 5/\auditrev{5} & 0/0 & 5/5 & 0/0 & 10/\auditrev{10} & --- \\
& \cfg{3,+3,-2,-2,-2} & 11/8 & 12/0 & 5/5 & 12/6 & 40/19 & --- \\
\end{tabular}
\end{ruledtabular}
\endgroup
\end{table*}
 
\section{Charge-flow consequences of irreducibility}
\label{sec:charge_flow}

The crystalline classification established above determines which primitive Weyl inventories can exist as standalone IWSM phases. We now ask how this bulk irreducibility constrains their boundary topology. The relevant invariant is not a unique microscopic Fermi-arc contour, which can vary with surface termination, energy, and surface potentials~\rev{\cite{hashimoto2017,devizorova2017,morali2019}}, but the net chiral spectral-flow incidence required by the projected bulk charges.

Consider a surface projection
\begin{equation}
	\pi:
	\mathbf{k}
	\longmapsto
	\bar{\mathbf{k}},
\end{equation}
from the bulk BZ to the surface BZ. If several bulk WPs project onto the same surface momentum $\bar{\mathbf{k}}_a$, the corresponding projected chiral charge is
\begin{equation}
	Q_a
	\equiv
	Q_{\rm proj}(\bar{\mathbf{k}}_a)
	=
	\sum_{j:\,\pi(\mathbf{k}_j)=\bar{\mathbf{k}}_a}
	C_j ,
	\label{eq:projected_charge}
\end{equation}
where $\mathbf{k}_j$ and $C_j$ denote the momentum and chiral charge of the $j$th bulk WP, respectively. Bulk--boundary correspondence requires the net chiral spectral-flow multiplicity incident on this projected terminal to satisfy~\rev{\cite{wan2011,fang2012multiweyl}}
\begin{equation}
	N_{{\rm ch},a}^{\rm net}
	=
	|Q_a|.
	\label{eq:projected_incidence}
\end{equation}
Equation~(\ref{eq:projected_incidence}) fixes the net oriented boundary incidence rather than the number of visibly separated branches on an arbitrary constant-energy contour. In particular, when several bulk nodes overlap in projection, only their summed charge $Q_a$ is topologically resolved. Surface termination or other boundary perturbations may reshape and reconnect individual Fermi-arc branches without changing this net incidence.

To expose the consequence of bulk irreducibility most directly, we first consider a resolved surface projection for which all member WPs of the IWSM project to distinct surface momenta. Let
\begin{equation}
	P_{+}
	=
	\{a_1,\ldots,a_r\},
	\qquad
	C(a_i)=p_i>0,
\end{equation}
and
\begin{equation}
	P_{-}
	=
	\{b_1,\ldots,b_s\},
	\qquad
	C(b_j)=-q_j<0,
\end{equation}
denote the positive- and negative-chirality projected terminals, respectively. Here $r$ and $s$ are the numbers of positive and negative terminals, while $p_i$ and $q_j$ are positive integers specifying their monopole-charge magnitudes. Charge neutrality gives
\begin{equation}
	\sum_{i=1}^{r}p_i
	=
	\sum_{j=1}^{s}q_j
	\equiv
	Q_{\rm tot}.
	\label{eq:terminal_neutrality}
\end{equation}

We characterize the topologically protected part of the boundary spectrum by removing topologically trivial closed surface contours and cancelling counterpropagating pairs of surface modes. The remaining energyresolved net chiral spectral flow can then be decomposed into oriented unit channels running from positive to negative projected charge. We define
\begin{equation}
	F_{ij}\in\mathbb{Z}_{\geq0}
	\label{eq:F_definition}
\end{equation}
as the number of such unit net-flow channels connecting the positive terminal $a_i$ to the negative terminal $b_j$. The resulting non-negative integer matrix
\begin{equation}
	F=(F_{ij})\in\mathbb{Z}_{\geq0}^{\,r\times s}
\end{equation}
will be referred to as the cycle-reduced chiral-incidence matrix. Bulk--boundary correspondence and charge conservation require
\begin{equation}
	\sum_{j=1}^{s}F_{ij}=p_i,
	\qquad
	\sum_{i=1}^{r}F_{ij}=q_j.
	\label{eq:flow_conservation}
\end{equation}
Thus the positive charges determine the row sums of $F$, the negative charges determine its column sums, and Eq.~(\ref{eq:terminal_neutrality}) guarantees equality of the total incoming and outgoing chiral flow.

Equation~(\ref{eq:flow_conservation}) immediately yields a boundary consequence of IWSM primitivity.

\paragraph{Connectedness of the irreducible boundary incidence.}
For a resolved projection of an IWSM, every cycle-reduced incidence matrix satisfying Eq.~(\ref{eq:flow_conservation}) defines a connected bipartite multigraph. To see this, suppose that the incidence graph were disconnected. Any connected component would contain a nonempty subset $I\subset\{1,\ldots,r\}$ of positive terminals and a nonempty subset $J\subset\{1,\ldots,s\}$ of negative terminals. Because no incidence channel leaves the component, Eq.~(\ref{eq:flow_conservation}) requires
\begin{equation}
	\sum_{i\in I}p_i
	=
	\sum_{j\in J}q_j.
	\label{eq:component_neutrality}
\end{equation}
The corresponding bulk nodes would therefore form a nonempty proper charge-neutral subinventory, contradicting the primitivity of the IWM. Hence the cycle-reduced boundary incidence of a resolved IWSM cannot factor into disconnected neutral blocks.

The converse does not hold. A reducible bulk Weyl inventory may acquire a connected surface network through boundary-induced mixing between otherwise independent neutral sectors. Consequently, connected incidence is a necessary but not sufficient boundary signature of an IWSM. By contrast, if the measured or calculated cycle-reduced incidence can be permuted into independent neutral blocks,
\begin{equation}
	F
	\sim
	F^{(1)}
	\oplus
	F^{(2)}
	\oplus
	\cdots
	\oplus
	F^{(n)},
	\qquad
	n>1,
	\label{eq:block_decomposition}
\end{equation}
with each block separately satisfying charge neutrality, the corresponding bulk inventory cannot represent one resolved IWSM. Bulk irreducibility must still be established independently from the complete signed inventory, exact MSG-orbit closure, and the common-band compatibility criteria introduced in \rev{Sec.~\ref{sec:irreducible_phase}}.

The local chiral charge fixes the valence of each terminal but not the global incidence architecture. For a two-band crossing on an $n$-fold rotation- or screw-invariant line, the leading symmetry-constrained Hamiltonian may be written as
\begin{equation}
	\mathcal{H}(\delta\mathbf{k})
	=
	v_z\delta k_z\sigma_z
	+
	f(k_+,k_-)\sigma_+
	+
	f^{*}(k_+,k_-)\sigma_-,
	\label{eq:kp_general}
\end{equation}
where
\begin{equation}
	k_{\pm}
	=
	\delta k_x\pm i\delta k_y,
	\qquad
	\sigma_{\pm}
	=
	\frac{\sigma_x\pm i\sigma_y}{2}.
\end{equation}
Here $\delta\mathbf{k}$ is measured from the crossing, $v_z$ is the velocity along the invariant axis, and the two basis states forming the crossing have rotation or screw eigenvalues $\lambda_c$ and $\lambda_v$. Symmetry requires~\cite{fang2012multiweyl,tsirkin2017composite}
\begin{equation}
	f
	\left(
	k_+e^{i2\pi/n},
	k_-e^{-i2\pi/n}
	\right)
	=
	\frac{\lambda_c}{\lambda_v}
	f(k_+,k_-).
	\label{eq:rotation_constraint}
\end{equation}
The lowest-order transverse monomial allowed by Eq.~(\ref{eq:rotation_constraint}) determines the local winding of the two-band Hamiltonian: linear, quadratic, and cubic transverse couplings produce stable WPs with $|C|=1$, $2$, and $3$, respectively. The charge-four realizations discussed below instead require the combined constraints of a chiral cubic little group~\rev{\cite{zhang2020quadruple,cui2021chargefour}}. The sign of $C$ is determined independently from the Berry-curvature flux or Wannier charge center winding. Thus crystalline representations determine the local terminal valence $|C|$, whereas the complete signed inventory and Eq.~(\ref{eq:flow_conservation}) determine how these terminals can be assembled globally.

\begin{figure*}[t]
\centering
\begingroup
\footnotesize
\setlength{\tabcolsep}{5pt}
\renewcommand{\arraystretch}{1.38}
\begin{tabular*}{0.98\textwidth}{@{\extracolsep{\fill}}lccc@{}}
\toprule
\textbf{Motif} &
\textbf{Pair} &
\textbf{Split} &
\textbf{Mixed} \\
\midrule
\raisebox{1.7em}{\textbf{Representative incidence}} &
\includegraphics[width=0.135\textwidth]{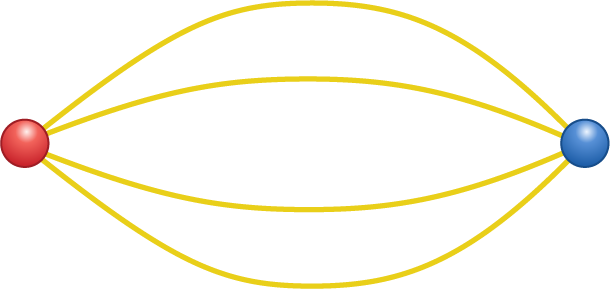} &
\includegraphics[width=0.135\textwidth]{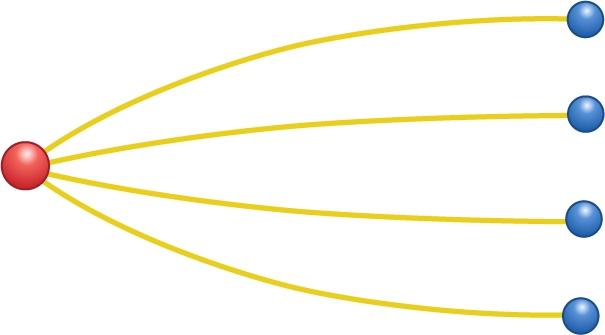} &
\includegraphics[width=0.135\textwidth]{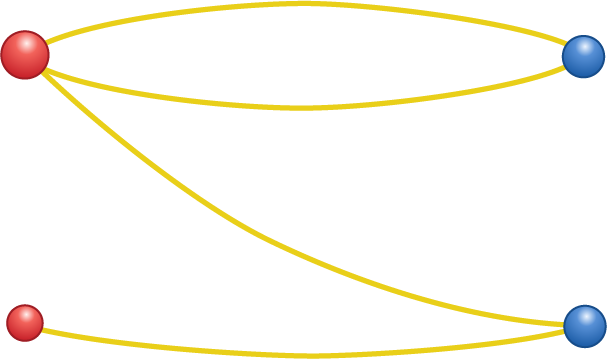} \\
\addlinespace[2pt]
\textbf{Terminal topology} &
one-to-one &
\shortstack{one-to-many\\or many-to-one} &
many-to-many \\
\midrule
\textbf{Parent-PG union$^{a}$} &
\shortstack{\{noncubic PGs except $C_1,C_i,C_s$\};\\\{$T,O$\}} &
$C_{3,4,6},D_{3,4,6},T,O$ &
$C_6,D_6$ \\
\addlinespace[2pt]
\textbf{Magnetic-space-group types$^{b}$} &
I--IV &
I--IV &
I--IV \\
\addlinespace[2pt]
\textbf{Cycle-reduced routing} &
unique &
unique &
multiple labeled \\
\bottomrule
\end{tabular*}
\endgroup
\caption{
	Charge-node architectures and symmetry domains of Pair, Split, and Mixed IWSMs.
	The representative configurations are $\{4,-4\}$, $\{4,-1,-1,-1,-1\}$, and $\{3,1,-2,-2\}$, respectively.
	Each yellow curve represents one unit of reduced net chiral incidence between positive (red) and negative (blue) terminals.
	For resolved projections, Pair and Split have unique scalar and vector incidences, whereas Mixed admits multiple labeled chiral-incidence matrices $F=(F_{ij})$ with fixed terminal charges [Eq.~(\ref{eq:mixed_example_1})].
	Superscripts $a$ and $b$ identify family unions of parent point groups (PGs) and MSG types, respectively, rather than symmetries common to every member.
	Configuration- and spin-resolved settings are given in \rev{Table~\ref{tab:iwsm-msg-reports}} and \rev{Tables~S16--S25} of the SM~\cite{supplemental}.
}
\label{fig:motifs}
\end{figure*}

This separation leads naturally to three IWSM charge-flow architectures [\rev{Fig.~\ref{fig:motifs}}]. Let $r$ and $s$ denote the numbers of resolved positive and negative terminals, respectively. We define
\begin{equation}
	\begin{cases}
		r=s=1, & \text{Pair},\\[2pt]
		\min(r,s)=1,\ \max(r,s)>1, & \text{Split},\\[2pt]
		r>1,\ s>1, & \text{Mixed}.
	\end{cases}
	\label{eq:motif_definition}
\end{equation}

For a Pair IWSM,
\begin{equation}
	X=\{q,-q\},
	\qquad
	q=1,2,3,4,
\end{equation}
the incidence matrix contains a single entry,
\begin{equation}
	F=(q).
	\label{eq:pair_matrix}
\end{equation}
The four Pair classes therefore share the same primitive one-to-one terminal topology while carrying one, two, three, or four units of net chiral spectral flow. Their local monopole valence changes with $q$, but their global incidence topology does not. \rev{The four inventory-labelled sectors of SM~\cite{supplemental} Fig.~S1 provide the corresponding energy-resolved Pair spectra.}

A Split IWSM has a single terminal in one chirality sector and multiple terminals in the other. For a representative positive source of charge $q$ compensated by negative terminals $-q_j$,
\begin{equation}
	q
	=
	\sum_{j=1}^{s}q_j,
\end{equation}
Eq.~(\ref{eq:flow_conservation}) uniquely fixes
\begin{equation}
	F
	=
	\begin{pmatrix}
		q_1 & q_2 & \cdots & q_s
	\end{pmatrix},
	\label{eq:split_matrix}
\end{equation}
with the transpose applying to the chirality-reversed many-to-one case. The four Split IWSMs are
\begin{equation}
	\begin{gathered}
	\{2,-1,-1\},\qquad \{3,-1,-1,-1\},\\
	\{4,-1,-1,-1,-1\},\qquad \{3,-2,-1\},
	\end{gathered}
	\label{eq:split_inventories}
\end{equation}
with corresponding cycle-reduced incidence vectors
\begin{equation}
	F_{\{2,-1,-1\}}
	=
	\begin{pmatrix}
		1 & 1
	\end{pmatrix},
	\qquad
	F_{\{3,-1,-1,-1\}}
	=
	\begin{pmatrix}
		1 & 1 & 1
	\end{pmatrix},
	\label{eq:split_vectors_1}
\end{equation}
and
\begin{equation}
	\begin{aligned}
	F_{\{4,-1,-1,-1,-1\}}
	&=
	\begin{pmatrix}
		1 & 1 & 1 & 1
	\end{pmatrix},\\[3pt]
		F_{\{3,-2,-1\}}
	&=
	\begin{pmatrix}
		2 & 1
	\end{pmatrix},
	\end{aligned}
	\label{eq:split_vectors_2}
\end{equation}
up to global chirality reversal. The first three classes realize equal-valence fan-out into unit-charge terminals, whereas $\{3,-2,-1\}$ realizes a heterovalent partition in which two units of net chiral flow terminate at the charge-$-2$ projection and one unit at the charge-$-1$ projection. The Split architecture is therefore defined by a one-to-many or many-to-one integer charge partition, rather than by the presence of unit-charge terminals alone. \rev{Energy-resolved boundary spectra for these realizations are provided in SM~\cite{supplemental} Sec.~IV, with the unit-terminal Split configurations in Fig.~S2 and the heterovalent configuration in Fig.~S3.}

For a Mixed IWSM, both chirality sectors contain multiple terminals and Eq.~(\ref{eq:flow_conservation}) fixes only the row and column sums of $F$. The internal labeled source--sink matching can therefore be nonunique. The two Mixed IWSM inventories are
\begin{equation}
	\{3,1,-2,-2\},
	\qquad
	\{3,3,-2,-2,-2\}.
	\label{eq:mixed_inventories}
\end{equation}

For $\{3,1,-2,-2\}$, the positive margins are $(3,1)$ and the negative margins are $(2,2)$. Two admissible cycle-reduced incidence matrices are
\begin{equation}
	\begin{aligned}
	F^{(a)}_{\{3,1,-2,-2\}}
	&=
	\begin{pmatrix}
		1 & 2\\
		1 & 0
	\end{pmatrix},\\[3pt]
		F^{(b)}_{\{3,1,-2,-2\}}
	&=
	\begin{pmatrix}
		2 & 1\\
		0 & 1
	\end{pmatrix}.
	\end{aligned}
	\label{eq:mixed_example_1}
\end{equation}
Both satisfy Eq.~(\ref{eq:flow_conservation}) and are connected, as required by primitivity. If the two charge-$-2$ terminals are regarded as unlabeled, the two matrices are related by their exchange; when the corresponding projections occupy distinct surface momenta, they represent different labeled boundary matchings.

The second Mixed class, $\{3,3,-2,-2,-2\}$, contains two positive charge-three terminals and three negative charge-two terminals. Its row margins are therefore $(3,3)$ and its column margins are $(2,2,2)$. Representative admissible incidence matrices include
\begin{equation}
	\begin{aligned}
	F^{(a)}_{\{3,3,-2,-2,-2\}}
	&=
	\begin{pmatrix}
		1 & 1 & 1\\
		1 & 1 & 1
	\end{pmatrix},\\[3pt]
		F^{(b)}_{\{3,3,-2,-2,-2\}}
	&=
	\begin{pmatrix}
		2 & 1 & 0\\
		0 & 1 & 2
	\end{pmatrix}.
	\end{aligned}
	\label{eq:mixed_example_2}
\end{equation}
Again, both matrices satisfy the same terminal margins and define connected incidence graphs. Up to permutations of symmetry-equivalent terminals carrying the same charge, they illustrate two distinct ways in which the six units of total chiral spectral flow can be distributed within the same primitive two-to-three charge-flow architecture. The bulk IWSM inventory fixes the terminal valences and the allowed integer margins, but it does not in general select a unique labeled internal routing.

Whether a particular boundary Hamiltonian provides a continuous switching path between distinct admissible Mixed matchings is not fixed by the bulk classification and remains a surface-specific question. Surface termination, surface potential, adsorption, or other boundary perturbations may select or reconfigure the realized matching while preserving the projected terminal charges and their net chiral incidences.

The Pair--Split--Mixed distinction therefore forms a hierarchy of boundary-routing freedom rather than a hierarchy of local Weyl charge. \rev{The point-group and MSG-type entries in Fig.~\ref{fig:motifs} describe unions over each family, not symmetries shared by every member.} Pair fixes one scalar incidence, Split fixes one row or column vector, and Mixed fixes only the integer margins of a generally nonunique matrix. This distinction is particularly clear among the five IWSM inventories containing a charge-three WP:
\begin{equation}
	\begin{gathered}
	\{3,-3\},\quad \{3,-1,-1,-1\},\quad \{3,-2,-1\},\\
	\{3,1,-2,-2\},\quad \{3,3,-2,-2,-2\}.
	\end{gathered}
\end{equation}
Although the local $|C|=3$ crossing can share the same leading cubic Weyl structure, its compensating charge sector places the complete phase in a Pair, Split, or Mixed architecture. The phase-level topology is therefore determined by the complete signed inventory rather than by the local triple-Weyl object alone.

The crystallographic support of these architectures is likewise inventory specific rather than monotonic in node number or motif complexity [\rev{Table~\ref{tab:iwsm-msg-reports}}]. \rev{For example, the three-terminal Split inventory $\{2,-1,-1\}$ is compatible with 166 BNS settings without SOC, more than the \tcnew{122} settings supporting the two-terminal $\{2,-2\}$ Pair.} Additional terminals can therefore complete a symmetry-required orbit or compatible band-connectivity pattern rather than simply imposing additional constraints. The number of WPs and the largest local $|C|$ are consequently not, by themselves, measures of the crystallographic accessibility of an IWSM.

Together, the local representation analysis and the global chiral-incidence matrix provide complementary information. The former determines the monopole valence carried by each Weyl terminal, whereas the latter encodes how the complete irreducible inventory distributes its topological spectral flow at a boundary. This charge-flow taxonomy supplies the physical organization underlying the explicit Pair, Split, and Mixed lattice realizations constructed in the following section.

\section{Constructive realizations of the three motifs}
\label{sec:constructive_realizations}

We now construct a symmetry-constrained lattice realization for each of the ten IWSM classes, thereby connecting the crystalline classification of \rev{Sec.~\ref{sec:irreducible_phase}} to the boundary charge-flow structures of \rev{Sec.~\ref{sec:charge_flow}}. The models are organized into the Pair, Split, and Mixed families shown in \rev{Figs.~\ref{fig:pair}--\ref{fig:mixed}}. Within each family, the symmetry representations of the crossing states determine the allowed local band couplings, the calculated Chern charges establish the signed Weyl-node configuration, and the surface spectra reveal its boundary manifestation. This construction separates two questions: which local Weyl crossings occur, and how their complete set forms one irreducible crystalline Weyl phase.

The lattice Hamiltonians are constructed using MSGCorep and MagneticTB from symmetry-compatible orbital bases and hopping terms~\rev{\cite{Liu2023MSGCorep,Zhang2022MagneticTB}}. All ten explicit models use single-valued representations without SOC; the additional double-valued settings in \rev{Table~\ref{tab:iwsm-msg-reports}} belong to the symmetry classification and are not separate model constructions presented here. For each model, we select the adjacent bands $N$ and $N+1$ forming the active Weyl sector and search their band separation $g_N(\mathbf{k})$ throughout the BZ. After local refinement, symmetry completion, and identification of reciprocal-lattice-equivalent momenta, no additional crossings of the selected pair are found within the numerical resolution of the search. The chiral charges are evaluated from the winding of hybrid Wannier charge centers (WCCs) on closed surfaces enclosing individual nodes, and semi-infinite surface spectra are calculated using the iterative Green-function method~\rev{\cite{gresch2017z2pack,wu2018wanniertools}}. The computational procedure, complete Hamiltonians, and node coordinates are provided in \rev{Secs.~II, III, and VI of the SM~\cite{supplemental}}, with the node data summarized in \rev{Table~S15}.

The supporting results follow the same inventory ordering throughout the SM~\cite{supplemental}. Section~I\,B and Tables~S4--S13 give the symmetry-screened settings; Sec.~V and Tables~S16--S25 list the final compatible WP channels. The unit-pair restriction is derived in Sec.~I\,A\,2. Sections~VI\,A--VI\,J specify the orbital bases and hopping Hamiltonians, with the perspective and top views in Figs.~S4--S13(a,b). Within these sections, the four Pair models precede the four Split models and the two Mixed models.

In the following, $\lambda(\rho)$ denotes the eigenvalue of the relevant rotation or screw operation for a crossing branch labeled by $\rho$. The transverse momenta $k_{\pm}$ and the corresponding selection rule are those defined in \rev{Sec.~\ref{sec:charge_flow}}. The quantity $E_{\rm cut}$ specifies the energy at which a surface spectral contour is evaluated in the plotting convention of the corresponding model. Because different WPs need not be isoenergetic, these contours are interpreted together with the energy-resolved surface spectra in \rev{Figs.~S1--S3} of the SM~\cite{supplemental}. The protected boundary information is the net chiral incidence associated with the projected charges, not a termination-independent contour at one energy.

\subsection{Pair IWSMs: a single pair with increasing chiral charge}
\label{sec:pair_models}

\begin{figure*}[!t]
	\centering
	\includegraphics[width=\textwidth]{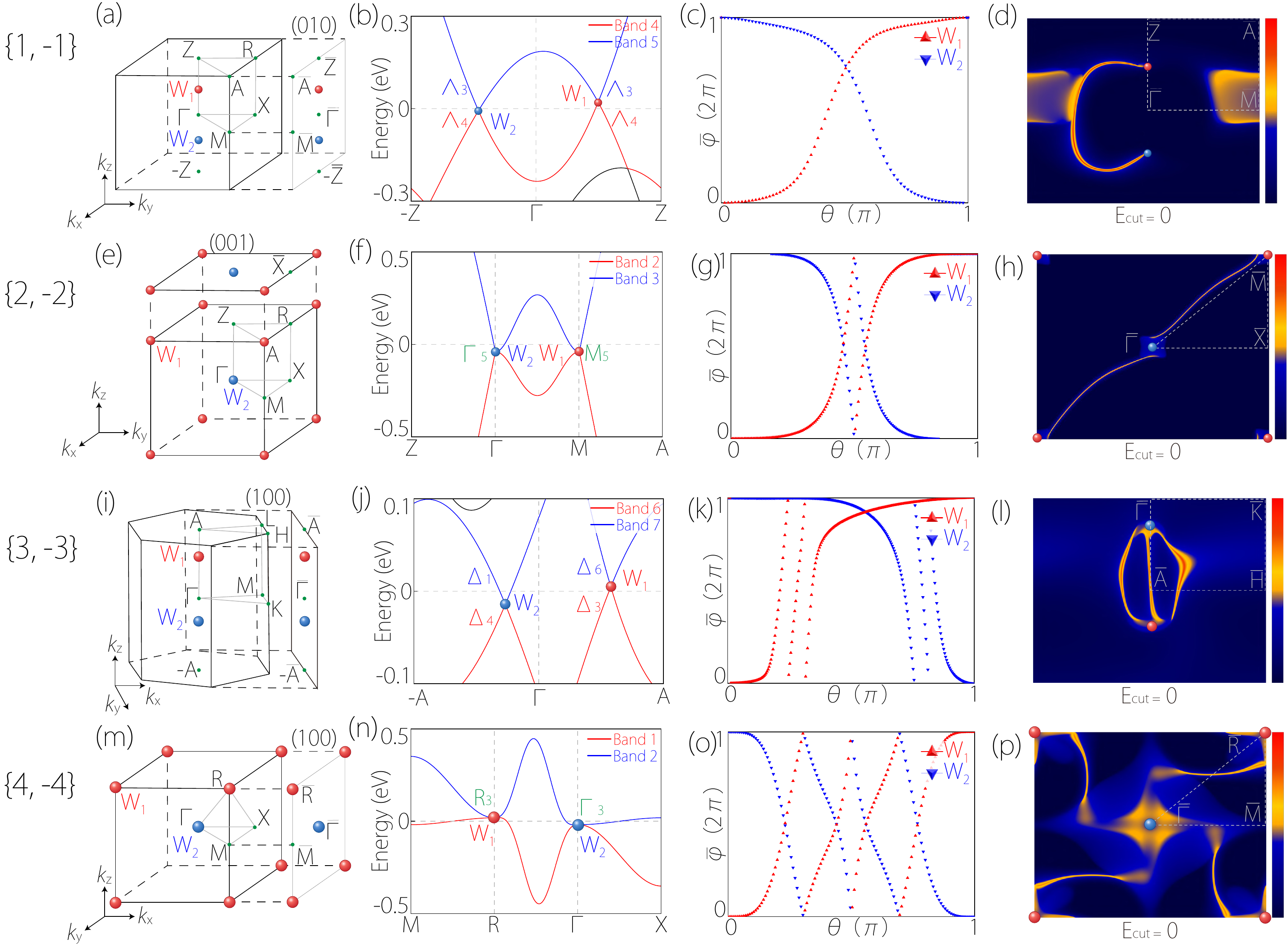}
	\caption{
		Symmetry-constrained lattice realizations of the four Pair IWSMs.
		(a)--(d) $\{1,-1\}$ in MSG 76.7.
		(e)--(h) $\{2,-2\}$ in MSG 91.103.
		(i)--(l) $\{3,-3\}$ in MSG 169.113.
		(m)--(p) $\{4,-4\}$ in MSG 207.40.
		All models use single-valued representations without spin-orbit coupling (SOC), with selected bands $(N,N+1)$ for $N=4,2,6,1$, respectively.
		Columns show bulk nodes and surface projections, symmetry-labeled band dispersions, Wannier charge center (WCC) winding, and surface spectral contours.
		Full-BZ scans of the band separation $g_N(\mathbf{k})=E_{N+1}(\mathbf{k})-E_N(\mathbf{k})$ confirm, within numerical resolution, that its only zeros are the nodes of the stated IWSM configuration in each model.
		The surfaces are $(010)$, $(001)$, $(100)$, and $(100)$, all at $E_{\rm cut}=0$.
		Spectral colors denote surface weight; WCC phases are plotted in units of $2\pi$ against the sphere polar angle $\theta$ in units of $\pi$.
	}
	\label{fig:pair}
\end{figure*}

The four Pair models realize $\{1,-1\}$, $\{2,-2\}$, $\{3,-3\}$, and $\{4,-4\}$ while retaining one positive and one negative Weyl node [\rev{Fig.~\ref{fig:pair}}]. They therefore provide a direct comparison in which the local chiral charge changes but the number of compensating nodes does not. For resolved projections, the incidence matrix remains $F=(q)$, with $q=1,2,3,4$, as established in Eq.~(\ref{eq:pair_matrix}).

Using MSG 76.7 as a representative symmetry setting, we construct a lattice model and select bands 4 and 5 as the active Weyl sector. Their two crossings on the screw axis through $\Gamma$ realize the unit-charge Pair configuration [\rev{Figs.~\ref{fig:pair}(a) and \ref{fig:pair}(b)}]. Both crossings involve the $\Lambda_4$ and $\Lambda_3$ branches, whose fourfold-screw eigenvalues satisfy
\begin{equation}
	\frac{\lambda(\Lambda_4)}{\lambda(\Lambda_3)}=i.
	\label{eq:model_pair_unit_ratio}
\end{equation}
The transverse momentum $k_+$ acquires the same phase under a fourfold rotation, so Eq.~(\ref{eq:rotation_constraint}) permits a linear interband coupling. The calculated WCC windings assign $C(W_1)=+1$ and $C(W_2)=-1$ [\rev{Fig.~\ref{fig:pair}(c)}], establishing the complete $\{1,-1\}$ configuration of the selected band pair. The $(010)$-surface contour at $E_{\rm cut}=0$ [\rev{Fig.~\ref{fig:pair}(d)}] is consistent with one unit of net chiral incidence at each separated projection.

For the double-Weyl Pair, we construct a symmetry-constrained lattice model using MSG 91.103 as a representative setting. Bands 2 and 3 of this model form the selected Weyl sector [\rev{Figs.~\ref{fig:pair}(e)--\ref{fig:pair}(h)}]. The two nodes occur at $M$ and $\Gamma$, in the $M_5$ and $\Gamma_5$ representation sectors, respectively \rev{[Figs.~\ref{fig:pair}(e,f)]}. Within each crossing doublet, the fourfold eigenvalues are $+i$ and $-i$, giving an eigenvalue ratio of $-1$. Linear transverse coupling is therefore forbidden, whereas quadratic terms proportional to $k_+^2$ or $k_-^2$ are allowed. The resulting nodes have quadratic transverse dispersion and linear axial dispersion. WCC winding gives $C(W_1)=+2$ at $M$ and $C(W_2)=-2$ at $\Gamma$ [\rev{Fig.~\ref{fig:pair}(g)}]. Their distinct projections on the $(001)$ surface each carry two units of net chiral incidence, with a representative contour shown at $E_{\rm cut}=0$ [\rev{Fig.~\ref{fig:pair}(h)}]. Relative to the unit-charge model, the local charge and boundary-channel multiplicity are doubled without changing the primitive Pair configuration.

To realize the triple-Weyl Pair, we construct a lattice Hamiltonian subject to the symmetry constraints of MSG 169.113 and select bands 6 and 7 as the active Weyl sector [\rev{Figs.~\ref{fig:pair}(i)--\ref{fig:pair}(l)}]. Two crossings occur on the sixfold screw axis through $\Gamma$ \rev{[Fig.~\ref{fig:pair}(i)]}: one between the $\Delta_4$ and $\Delta_1$ branches, and the other between the $\Delta_6$ and $\Delta_3$ branches. Both pairs have a screw-eigenvalue ratio of $-1$ \rev{[Fig.~\ref{fig:pair}(j)]}. Under a sixfold rotation, this phase is carried by $k_{\pm}^3$, whereas linear and quadratic transverse terms are forbidden. The crossings therefore realize the symmetry-allowed cubic transverse structure of triple-Weyl points. Opposite threefold WCC windings establish $C(W_1)=+3$ and $C(W_2)=-3$ [\rev{Fig.~\ref{fig:pair}(k)}]. The $(100)$-surface spectrum at $E_{\rm cut}=0$ [\rev{Fig.~\ref{fig:pair}(l)}] provides the corresponding three-channel Pair realization. The combination of the screw representations and the enclosing-surface charge calculation distinguishes each triple-Weyl point from a cluster of unresolved unit-charge nodes.

For the charge-four Pair, we construct a symmetry-constrained lattice model based on MSG 207.40. Bands 1 and 2 of this model form the active Weyl sector, with nodes in the $R_3$ and $\Gamma_3$ sectors \rev{[Figs.~\ref{fig:pair}(m,n)]}. Unlike the axial double- and triple-Weyl constructions, the charge-four crossing is constrained by the combined operations of the chiral cubic little group. Its topology cannot be inferred from the eigenvalue mismatch of one rotation alone. The explicit Hamiltonian in \rev{Sec.~VI\,D of the SM~\cite{supplemental}}, together with the fourfold WCC winding, establishes $C(W_1)=+4$ at $R$ and $C(W_2)=-4$ at $\Gamma$ [\rev{Fig.~\ref{fig:pair}(o)}]. The $(100)$-surface contour at $E_{\rm cut}=0$ [\rev{Fig.~\ref{fig:pair}(p)}] is a representative boundary realization of the four-channel Pair configuration.

These four models demonstrate that increasing the chiral charge of the constituent nodes does not require increasing the number of nodes in an irreducible Weyl complex. The local structures change from linear through quadratic and cubic to the chiral-cubic charge-four crossing, while the complete signed inventory retains the Pair form. The integer $q$ determines the net boundary multiplicity, but the unique source--sink assignment follows from the complete two-node configuration rather than from the local dispersion.

\subsection{Split IWSMs: compensation by multiple Weyl nodes}
\label{sec:split_models}

\begin{figure*}[!t]
	\centering
	\includegraphics[width=\textwidth]{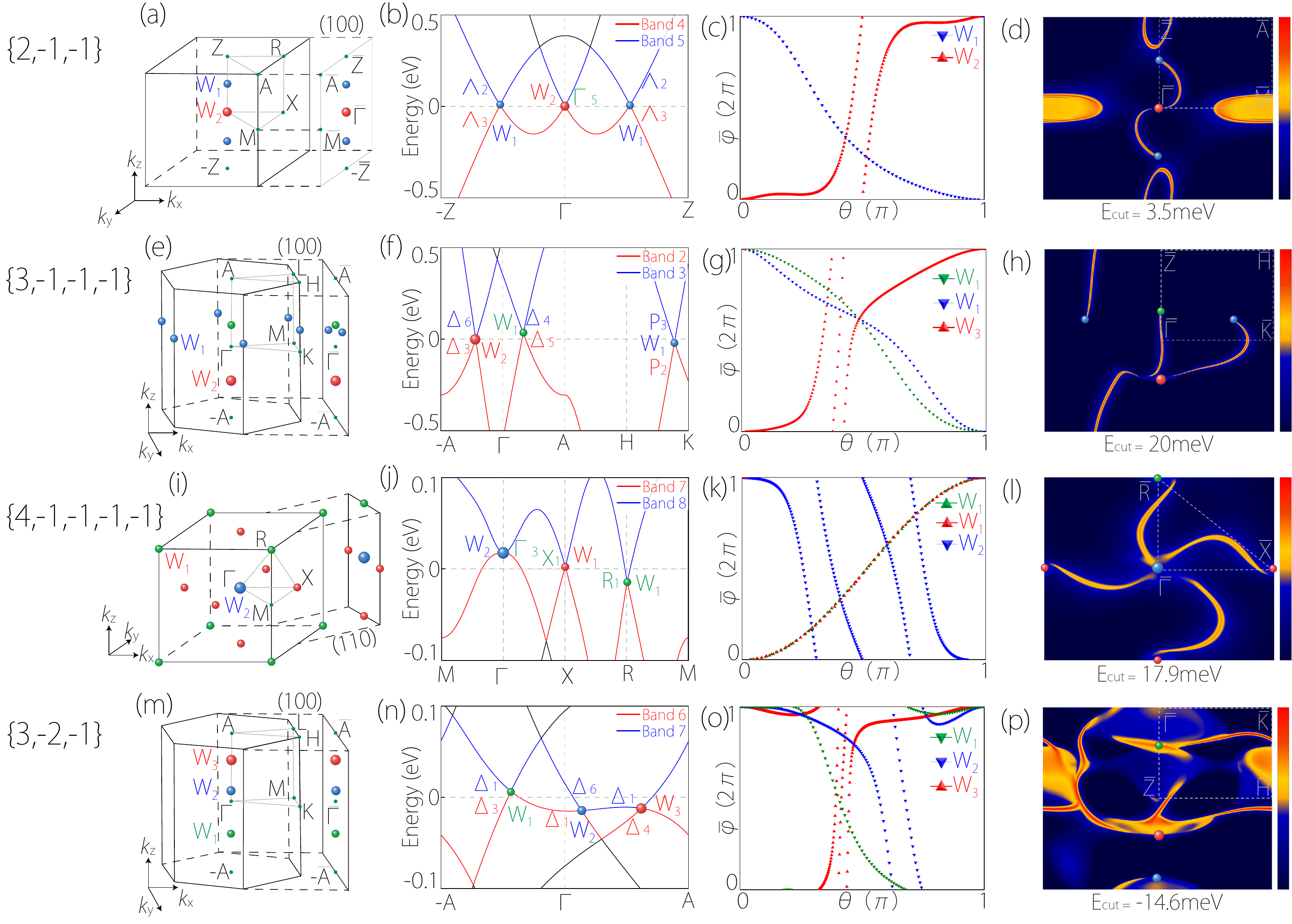}
	\caption{
		Symmetry-constrained lattice realizations of the four Split IWSMs.
		(a)--(d) $\{2,-1,-1\}$ in MSG 95.141.
		(e)--(h) $\{3,-1,-1,-1\}$ in MSG 169.113.
		(i)--(l) $\{4,-1,-1,-1,-1\}$ in MSG 212.59.
		(m)--(p) $\{3,-2,-1\}$ in MSG 178.159.
		All models use single-valued representations without SOC, with selected bands $(N,N+1)$ for $N=4,2,7,6$, respectively.
		Columns show bulk nodes and projections, symmetry-labeled bands, WCC winding, and surface spectral contours.
		Full-BZ scans of $g_N(\mathbf{k})=E_{N+1}(\mathbf{k})-E_N(\mathbf{k})$ confirm, within numerical resolution, that the selected band pair is degenerate only at the stated IWSM nodes.
		The numerical inventory in (i)--(l) is $\{-4,1,1,1,1\}$, globally reversed relative to the canonical class label.
		The surfaces are $(100)$, $(100)$, $(1\bar{1}0)$, and $(100)$ at $E_{\rm cut}=3.5$, $20$, $17.9$, and $-14.6\,\mathrm{meV}$, respectively.
		Spectral colors denote surface weight; node and WCC colors follow the labels within each row.
		Coincident projections carry summed chiral charges.
	}
	\label{fig:split}
\end{figure*}

The four Split models realize
$\{2,-1,-1\}$,
$\{3,-1,-1,-1\}$,
$\{4,-1,-1,-1,-1\}$, and
$\{3,-2,-1\}$
[\rev{Fig.~\ref{fig:split}}]. In each case, one higher-charge node is compensated by several oppositely charged nodes within the same selected adjacent-band pair. The first three configurations have unit-charge compensating nodes, whereas the last combines charge-one and charge-two nodes. Their resolved boundary incidences follow the vectors given in Eqs.~(\ref{eq:split_vectors_1}) and (\ref{eq:split_vectors_2}).

Using MSG 95.141 as a representative symmetry setting, we construct a lattice model whose bands 4 and 5 realize the $\{2,-1,-1\}$ Split configuration [\rev{Figs.~\ref{fig:split}(a)--\ref{fig:split}(d)}]. A double-Weyl point occurs at $\Gamma$ in the $\Gamma_5$ sector, accompanied by two symmetry-related unit-charge WPs on the screw axis \rev{[Figs.~\ref{fig:split}(a,b)]}. The fourfold eigenvalue ratio at the central crossing is $-1$, allowing quadratic but not linear transverse coupling. The off-center $\Lambda_3$--$\Lambda_2$ crossings instead have
$\lambda(\Lambda_3)/\lambda(\Lambda_2)=i$,
which permits linear coupling. WCC winding assigns $C(W_2)=+2$ to the central node and $C(W_1)=-1$ to each off-center node [\rev{Fig.~\ref{fig:split}(c)}]. The selected band pair thus realizes exactly one $\{2,-1,-1\}$ IWM. Its $(100)$-surface contour at $E_{\rm cut}=3.5\,\mathrm{meV}$ [\rev{Fig.~\ref{fig:split}(d)}] displays the boundary structure associated with one charge-two projection and two unit-charge projections. In the resolved incidence description, the two net channels are assigned to different compensating nodes, giving $F=(1,1)$.

For the $\{3,-1,-1,-1\}$ Split configuration, we construct a symmetry-constrained lattice model using MSG 169.113 as a representative setting [\rev{Figs.~\ref{fig:split}(e)--\ref{fig:split}(h)}]. In this model, bands 2 and 3 realize a triple-Weyl node compensated by three unit-charge nodes. The high-charge crossing involves the $\Delta_3$ and $\Delta_6$ branches, with
$\lambda(\Delta_6)/\lambda(\Delta_3)=-1$,
and consequently has cubic transverse coupling. The unit-charge crossings involve the $\Delta_5$--$\Delta_4$ branches on the principal screw axis and the $P_2$--$P_3$ branches on the $C_3$-invariant high-symmetry line $H$--$K$. Their eigenvalue ratios are $e^{i\pi/3}$ for the sixfold operation and $e^{i2\pi/3}$ for the threefold operation, respectively; both permit linear transverse coupling in their corresponding little groups. The three unit-charge nodes comprise one axial node and a symmetry-related pair, rather than a single three-member orbit \rev{[Figs.~\ref{fig:split}(e,f)]}. WCC winding gives one $C=+3$ node and three $C=-1$ nodes [\rev{Fig.~\ref{fig:split}(g)}], yielding the irreducible configuration $\{3,-1,-1,-1\}$. The $(100)$-surface contour at $E_{\rm cut}=20\,\mathrm{meV}$ [\rev{Fig.~\ref{fig:split}(h)}] is consistent with the resolved incidence vector $F=(1,1,1)$.

For the charge-four Split configuration, we construct a lattice model using MSG 212.59 as a representative symmetry setting and select bands 7 and 8 as the active Weyl sector [\rev{Figs.~\ref{fig:split}(i)--\ref{fig:split}(l)}]. The high-charge node lies at $\Gamma$ in the $\Gamma_3$ sector and is protected by the chiral cubic little group. Its compensating nodes comprise three symmetry-related $X$-type WPs and one WP at $R$, corresponding to the $X_1$ and $R_1$ sectors. Thus the four unit-charge nodes are supplied by a three-member orbit and a separate one-member orbit \rev{[Figs.~\ref{fig:split}(i,j)]}. The calculated charges are $C(W_2)=-4$ and $C(W_1)=+1$ for each compensating node [\rev{Fig.~\ref{fig:split}(k)}]. The numerical configuration is therefore $\{-4,1,1,1,1\}$, which belongs to the canonical $\{4,-1,-1,-1,-1\}$ class under global chirality reversal.

For a projection that resolves all five bulk nodes, this numerical representative has the many-to-one incidence $F=(1,1,1,1)^{\mathsf T}$. The actual $(1\bar{1}0)$-surface contour shown at $E_{\rm cut}=17.9\,\mathrm{meV}$ [\rev{Fig.~\ref{fig:split}(l)}] must instead be interpreted with the projection rule in Eq.~(\ref{eq:projected_charge}): bulk nodes that coincide in the surface BZ contribute their summed chiral charge. The bulk five-node configuration is unchanged by such projection overlap, but the number of distinguishable surface terminals need not equal the number of bulk WPs. This model consequently demonstrates both the charge-four Split realization and the need to distinguish a complete bulk inventory from its surface projection.

To realize the heterovalent Split configuration $\{3,-2,-1\}$, we construct a symmetry-constrained lattice model based on MSG 178.159. Bands 6 and 7 of this model exhibit the three constituent Weyl crossings [\rev{Figs.~\ref{fig:split}(m)--\ref{fig:split}(p)}]. All three nodes lie on the sixfold screw axis through $\Gamma$ \rev{[Fig.~\ref{fig:split}(m)]}, but their crossing branches have different symmetry eigenvalue mismatches \rev{[Fig.~\ref{fig:split}(n)]}. The $\Delta_3$--$\Delta_1$ crossing has ratio
$\lambda(\Delta_3)/\lambda(\Delta_1)=e^{i\pi/3}$
and permits linear transverse coupling. The $\Delta_1$--$\Delta_6$ crossing has ratio
$\lambda(\Delta_6)/\lambda(\Delta_1)=e^{-i2\pi/3}$
and permits quadratic coupling, while the $\Delta_4$--$\Delta_1$ crossing has ratio $-1$ and permits cubic coupling. WCC winding \rev{[Fig.~\ref{fig:split}(o)]} establishes
\begin{equation}
	C(W_1)=-1,\qquad
	C(W_2)=-2,\qquad
	C(W_3)=+3.
	\label{eq:model_heterovalent_charges}
\end{equation}
The complete selected-band configuration is therefore one $\{3,-2,-1\}$ IWM, not a superposition of independent charge-one, charge-two, and charge-three pairs. The $(100)$-surface contour at $E_{\rm cut}=-14.6\,\mathrm{meV}$ [\rev{Fig.~\ref{fig:split}(p)}] illustrates its asymmetric boundary incidence: the charge-three node supplies two net channels to the charge-two terminal and one to the unit-charge terminal in the resolved description, giving $F=(2,1)$.

Comparison of $\{3,-1,-1,-1\}$ with $\{3,-2,-1\}$ isolates the role of the compensating Weyl sector at fixed maximum charge. Both contain a triple-Weyl node, but one is completed by three unit-charge nodes and the other by a double-Weyl node and a unit-charge node. Their distinct incidence vectors are fixed by the complete signed configurations. These integers specify net chiral-channel multiplicities; they do not determine electrical-current or wave-power ratios.

\subsection{Mixed IWSMs: multiple nodes in both chirality sectors}
\label{sec:mixed_models}

\begin{figure*}[!t]
	\centering
	\includegraphics[width=\textwidth]{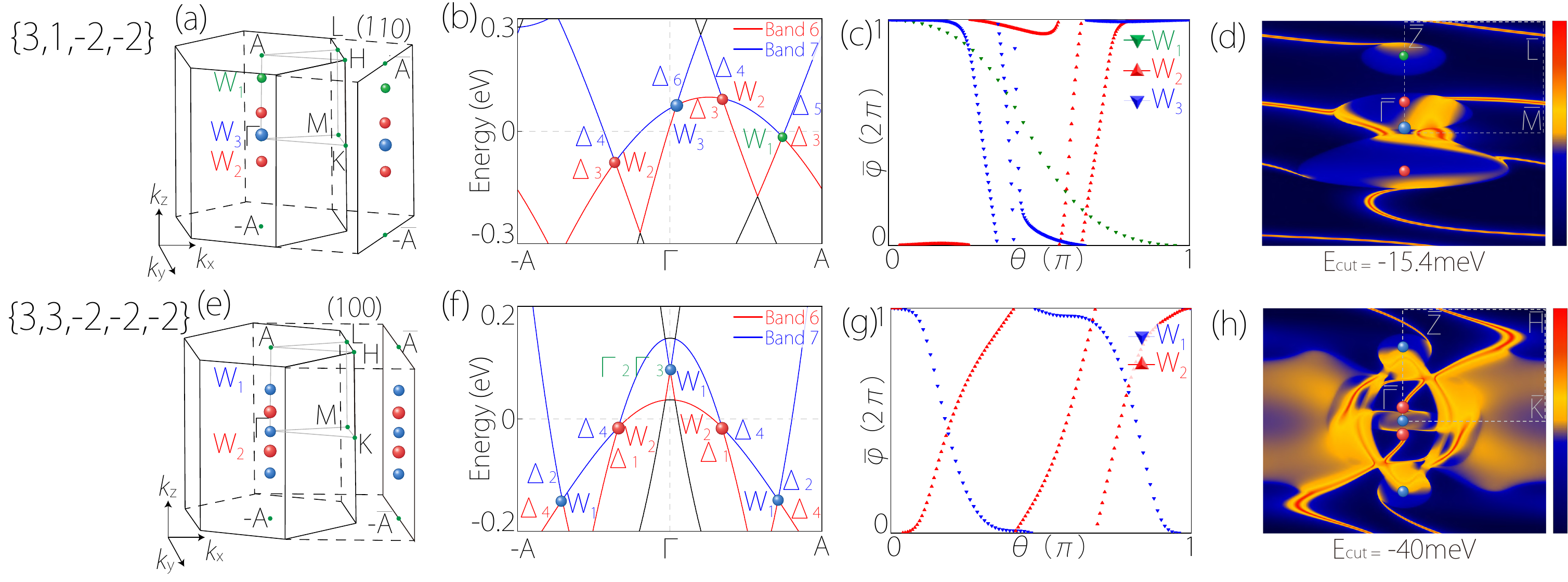}
	\caption{
		Lattice realizations of the two Mixed IWSMs.
		(a)--(d) MSG 169.113 realizes $\{-3,-1,2,2\}$, equivalent to $\{3,1,-2,-2\}$ under global chirality reversal.
		(e)--(h) Type-II MSG 169.114 realizes $\{3,3,-2,-2,-2\}$.
		Both models use single-valued representations without SOC and select bands 6 and 7.
		Columns show bulk nodes and projections, symmetry-labeled bands, WCC winding, and surface spectral contours.
		Full-BZ scans of $g_6(\mathbf{k})=E_7(\mathbf{k})-E_6(\mathbf{k})$ confirm, within numerical resolution, that its complete zero set contains exactly the four- or five-node IWSM configuration, respectively.
		The labels in (f) identify unitary-subgroup representations; magnetic corepresentations are listed in \rev{Table~S25 of the SM~\cite{supplemental}}.
		The surfaces are $(110)$ and $(100)$ at $E_{\rm cut}=-15.4$ and $-40\,\mathrm{meV}$.
		Spectral colors denote surface weight.
		The contours show representative boundary realizations, not switching between the admissible labeled incidence matrices $F=(F_{ij})$ illustrated in Eqs.~(\ref{eq:mixed_example_1}) and (\ref{eq:mixed_example_2}).
	}
	\label{fig:mixed}
\end{figure*}

The two Mixed models realize $\{3,1,-2,-2\}$ and $\{3,3,-2,-2,-2\}$ [\rev{Fig.~\ref{fig:mixed}}]. Both have more than one node in each chirality sector, so the bulk charges fix the row and column sums of the incidence matrix without generally selecting its individual entries. The models establish crystalline realizations of these configurations and their surface spectra; they do not by themselves demonstrate switching between every admissible matrix listed in \rev{Sec.~\ref{sec:charge_flow}}.

Taking MSG 169.113 as a representative symmetry setting, we construct a lattice Hamiltonian subject to its symmetry constraints. In this model, we select bands 6 and 7, whose crossings on the screw axis through $\Gamma$ realize the four-node Mixed configuration [\rev{Figs.~\ref{fig:mixed}(a)--\ref{fig:mixed}(d)}]. Two $\Delta_3$--$\Delta_4$ crossings have
$\lambda(\Delta_4)/\lambda(\Delta_3)=e^{i2\pi/3}$
and realize quadratic transverse coupling. A $\Delta_3$--$\Delta_6$ crossing has ratio $-1$ and realizes cubic coupling, whereas the $\Delta_3$--$\Delta_5$ crossing has ratio
$\lambda(\Delta_5)/\lambda(\Delta_3)=e^{i\pi/3}$
and permits linear coupling. Thus the same selected band pair contains one unit-charge node, two double-Weyl nodes, and one triple-Weyl node \rev{[Figs.~\ref{fig:mixed}(a,b)]}.

The WCC windings \rev{[Fig.~\ref{fig:mixed}(c)]} determine the numerical charges as
\begin{equation}
	\begin{gathered}
	C(W_1)=-1,\qquad C(W_{2a})=C(W_{2b})=+2,\\
	C(W_3)=-3,
	\end{gathered}
	\label{eq:model_mixed_four_charges}
\end{equation}
where the subscripts $a$ and $b$ distinguish the two symmetry-inequivalent double-Weyl nodes listed in \rev{Table~S15 of the SM~\cite{supplemental}}. The resulting $\{-3,-1,2,2\}$ inventory is the global chirality reversal of the canonical $\{3,1,-2,-2\}$ class. For the numerical representative, the positive and negative charge magnitudes are $(2,2)$ and $(3,1)$, respectively; after global chirality reversal, these become the canonical margins used in Eq.~(\ref{eq:mixed_example_1}). The $(110)$-surface contour at $E_{\rm cut}=-15.4\,\mathrm{meV}$ [\rev{Fig.~\ref{fig:mixed}(d)}] provides one boundary realization consistent with the four terminal charges. Their net incidences are fixed, but the bulk classification does not select a unique labeled pairing of the surface branches.

For the five-node Mixed configuration $\{3,3,-2,-2,-2\}$, we construct a lattice Hamiltonian using single-valued representations of the Type-II MSG 169.114 [\rev{Figs.~\ref{fig:mixed}(e)--\ref{fig:mixed}(h)}]. Selecting bands 6 and 7 of this model yields two triple-Weyl nodes and three double-Weyl nodes on the sixfold screw axis \rev{[Fig.~\ref{fig:mixed}(e)]}. The double-Weyl node at $\Gamma$ belongs to the paired $\Gamma_2$--$\Gamma_3$ sector. The labels shown in \rev{Fig.~\ref{fig:mixed}(f)} identify the constituent representations of the unitary subgroup; the corresponding magnetic corepresentations are listed in \rev{Table~S25} of the SM~\cite{supplemental}. Their sixfold eigenvalue ratio permits quadratic transverse coupling. The two off-center double-Weyl nodes arise from $\Delta_2$--$\Delta_4$ crossings with
$\lambda(\Delta_4)/\lambda(\Delta_2)=e^{-i2\pi/3}$,
whereas the $\Delta_1$--$\Delta_4$ crossings have eigenvalue ratio $-1$ and produce the two triple-Weyl nodes.

WCC winding assigns $C=-2$ to the central node and its two off-center double-Weyl counterparts, and $C=+3$ to each triple-Weyl node [\rev{Fig.~\ref{fig:mixed}(g)}]. The complete configuration contains no unit-charge WPs. Its resolved positive and negative margins are $(3,3)$ and $(2,2,2)$, respectively, giving a two-to-three incidence matrix with six units of total net chiral flow. The two matrices in Eq.~(\ref{eq:mixed_example_2}) describe distinct charge-conserving incidence patterns compatible with these margins. The $(100)$-surface contour at $E_{\rm cut}=-40\,\mathrm{meV}$ [\rev{Fig.~\ref{fig:mixed}(h)}], together with \rev{the corresponding inventory-labelled sector of SM~\cite{supplemental} Fig.~S3}, supplies a concrete surface realization of this Mixed configuration. It does not imply that both admissible incidence patterns are realized by the same boundary Hamiltonian.

The ten constructions collectively establish a lattice realization for every admitted IWSM class within the stated single-valued model scope. Their comparison shows that the local Weyl charge and the complete crystalline configuration encode different information. Pair models increase the chiral charge while retaining two nodes; Split models distribute the compensating charge among several nodes; and Mixed models place multiple high-charge nodes in both chirality sectors. In every case, the complete selected-band inventory identifies the IWSM, while the boundary spectra test the projected chiral incidence. Detailed Fermi-arc contours remain properties of the chosen surface and energy, rather than additional bulk classification data.

\section{Material and functional research landscape}
\label{sec:material_functional_landscape}

The classification and lattice constructions establish which primitive Weyl configurations admit crystalline realizations. Identifying these configurations in a material requires a further distinction between symmetry-allowed band crossings and the electronic or phonon spectrum realized under the relevant physical conditions. We therefore distinguish published material predictions and measurements, established mechanisms in related topological systems, and IWSM-specific tests motivated by the present classification \rev{[Figs.~\ref{fig:pair-applications}--\ref{fig:landscape}]}. The literature platforms discussed below provide candidate realizations or physical precedents; their identification as IWSMs additionally requires the complete Weyl-node configuration of the selected adjacent-band pair, its signed charges, and consistency with the actual crystal and magnetic symmetries.

\rev{Tables~S16--S25} of the SM~\cite{supplemental} provide the configuration-resolved MSG and irrep/corepresentation conditions for this comparison. The single-valued and double-valued entries must be used separately. In particular, a nonmagnetic crystal with crystallographic space group $G$ is compared with the corresponding Type-II grey MSG, including time reversal, rather than with the Type-I group obtained by omitting time reversal. For magnetic electronic candidates, the relevant MSG must instead be determined from both the crystal structure and the specified magnetic order. Agreement of a charge configuration or a crystallographic space-group number alone is not sufficient to establish a complete IWSM realization.

\subsection{Pair: material candidates, charge spectroscopy, and wave control}
\label{sec:pair_functions}

Magnetic exchange provides an established theoretical route to the unit-charge Pair configuration. Wang \textit{et al.} predicted a single pair of WPs in spin-aligned EuCd$_2$As$_2$, and Soh \textit{et al.} investigated the exchange-induced Weyl scenario in this material~\rev{\cite{wang2019eucd2as2,soh2019ideal}}. For moments aligned along the crystallographic $c$ axis, the relevant magnetic symmetry is $P\bar{3}m'1$ (BNS No.~164.89), rather than the nonmagnetic parent space group $P\bar{3}m1$ (No.~164)~\cite{Valadkhani2023}. Its experimental electronic structure nevertheless requires caution: Santos-Cottin \textit{et al.} reported a magnetic semiconducting state using transport, optical spectroscopy, and excited-state photoemission spectroscopy~\cite{SantosCottin2023}. EuCd$_2$As$_2$ is therefore treated here as a condition-dependent candidate rather than an unqualified experimental realization of a Pair IWSM.

A related first-principles route is the ferromagnetic configuration of MnBi$_2$Te$_4$, for which a single-pair Weyl phase was predicted~\rev{\cite{li2019mnbi2te4}}. The corresponding $c$-axis ferromagnetic symmetry is $R\bar{3}m'$ (BNS No.~166.101), distinct from the symmetry of its antiferromagnetic phase~\cite{Bernevig2022Magnetic}. Gao \textit{et al.} further predicted an intrinsic ferromagnetic WSM with a single pair of WPs in MnSn$_2$Sb$_2$Te$_6$-B~\cite{Gao2023MnX2B2T6}. In that work, the calculated $R\bar{3}m$ crystal structure and out-of-plane ferromagnetic order correspond to the same $R\bar{3}m'$ magnetic symmetry. The single-pair prediction concerns MnSn$_2$Sb$_2$Te$_6$-B specifically, rather than every member of the investigated family, which also contains ferromagnetic axion-insulator candidates. These magnetic systems are included as literature precedents for the $\{1,-1\}$ configuration; their assignment under the present criterion requires a full-MSG and selected-band comparison with \rev{Table~S16 of the SM~\cite{supplemental}}, not merely agreement with the reported two-node count.

For the $\{2,-2\}$ Pair class, first-principles calculations predict single pairs of charge-two Weyl fermions in the nonmagnetic boron allotropes HDSBC-B$_{20}$ and CR-B$_{12}$ in the spinless limit~\rev{\cite{zheng2026boron}}. The two HDSBC-B$_{20}$ enantiomers belong to $P4_3$ (No.~78) and $P4_1$ (No.~76), corresponding to grey MSGs 78.20 and 76.8, respectively. CR-B$_{12}$ belongs to $R32$ (No.~155), corresponding to grey MSG 155.46. All three settings occur in the single-valued Type-II entries of \rev{Table~S17 of the SM~\cite{supplemental}}. The agreement is therefore specific to the spinless electronic description; it should not be interpreted as establishing the same exact two-node configuration after SOC is restored.

An additional electronic candidate is the all-$sp^2$ carbon allotrope HSRDC-C$_{28}$ proposed by Bai \textit{et al.}~\rev{\cite{bai2026sp2carbon}}. Its calculated band structure contains a single pair of charge-two WPs protected by fourfold rotation and time-reversal symmetry. The crystal belongs to $I4_1$ (No.~80), whose nonmagnetic grey MSG 80.30 is likewise included in the single-valued Type-II sector of \rev{Table~S17 of the SM~\cite{supplemental}}. Together, these boron and carbon allotropes provide material-level examples consistent with the symmetry settings identified for the spinless $\{2,-2\}$ class. Detailed comparisons of Weyl-point locations and representation labels must use the same reciprocal-cell and symmetry conventions as the catalogue.

For the $\{4,-4\}$ architecture, Yang \textit{et al.} predicted a single pair of charge-four Weyl phonons in $P23$-type BeH$_2$~\rev{\cite{yang2023beh2}}. The crystallographic space group $P23$ (No.~195), together with time reversal in the harmonic phonon problem, corresponds to grey MSG 195.2. This setting appears in the single-valued Type-II entries of \rev{Table~S19 of the SM~\cite{supplemental}}. BeH$_2$ thus provides a candidate topological phonon material for the charge-four Pair configuration, rather than an electronic Weyl semimetal. Among the representative material reports considered here, no independently established standalone $\{3,-3\}$ material realization is identified.

\begin{figure}[!t]
	\centering
	\includegraphics[width=\columnwidth]{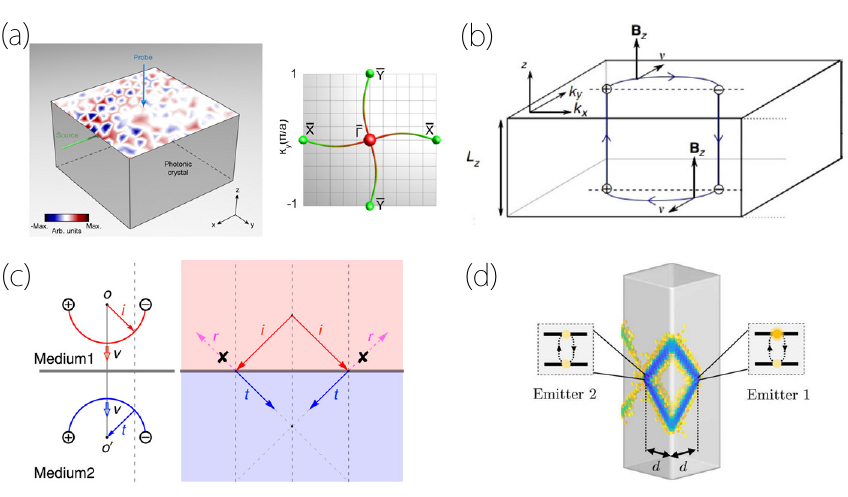}
	\caption{
		Published precedents for Pair-IWSM spectroscopy and wave control.
		(a) Source--probe and connectivity schematics from a charge-four photonic Weyl experiment.
		(b) Semiclassical Weyl orbit combining surface Fermi arcs and bulk chiral Landau channels.
		(c) Boundary-engineered reflectionless negative refraction in a theoretical two-WP photonic medium.
		(d) Proposed Fermi-arc-mediated coupling of quantum emitters on adjacent facets.
		Panels (a)--(d) are adapted from Refs.~\rev{\cite{chen2022chargefour,zhang2016generic,liu2022reflectionless,garciaelcano2023quantumlink}}, respectively.
		These studies establish measurement and physical-mechanism precedents, not independently certified Pair-IWSM devices.
	}
	\label{fig:pair-applications}
\end{figure}

Charge-resolved surface spectroscopy is a direct route for testing the boundary multiplicity of a Pair configuration. Chen \textit{et al.} experimentally reconstructed a charge-four WP and its quadruple-helicoid surface dispersion in a three-dimensional photonic crystal~\rev{\cite{chen2022chargefour}} [\rev{Fig.~\ref{fig:pair-applications}(a)}]. This establishes a method for resolving high-charge spectral flow, but it does not by itself establish the complete $\{4,-4\}$ inventory of a selected band pair. Applied to a candidate Pair IWSM, the additional test would be to establish the complete bulk pair and follow the net surface spectral flow associated with its two resolved projections. Within the assumptions of \rev{Sec.~\ref{sec:charge_flow}}, the corresponding incidence is $F=(q)$ for $\{q,-q\}$. This statement concerns the net chiral multiplicity, not the visibility of $q$ separate arcs on every constant-energy contour.

Surface--bulk Weyl orbits provide a complementary transport mechanism. Theories of Fermi-arc quantum oscillations describe closed trajectories that combine propagation along opposite surfaces with bulk chiral Landau channels~\rev{\cite{potter2014,zhang2016generic}} [\rev{Fig.~\ref{fig:pair-applications}(b)}]. Transport measurements in the Dirac semimetal Cd$_3$As$_2$ provided evidence consistent with this surface--bulk mechanism~\rev{\cite{moll2016}}. These results motivate Weyl-orbit measurements in a verified Pair IWSM, where the active Weyl sector contains only one opposite-charge node pair. They do not imply that a charge-$q$ pair produces $q$ independently resolved oscillation frequencies: the orbit quantization also depends on the surface dispersion, chemical potential, field orientation, and sample geometry~\rev{\cite{potter2014,zhang2016generic}}.

Optical charge readout likewise requires conditions beyond minimality. The theory of the quantized circular photogalvanic effect identifies a regime in which the trace of the injection response is governed by the active Weyl charge, provided that the crystal symmetry and optical frequency window allow a non-cancelling contribution from the relevant nodes~\rev{\cite{dejuan2017cpge}}. A Pair inventory alone does not establish this regime. In particular, the contributions of opposite chiralities, the allowed optical transitions, and the participation of additional bands must be evaluated for the chosen material. The classification specifies the available Weyl charges, whereas the response calculation determines whether they can be resolved optically.

Fermi-arc refraction offers a related wave-control route. Liu \textit{et al.} theoretically demonstrated all-angle reflectionless negative refraction in an ideal two-WP photonic medium with appropriately engineered boundary conditions~\rev{\cite{liu2022reflectionless}} [\rev{Fig.~\ref{fig:pair-applications}(c)}]. Independently, He \textit{et al.} experimentally observed topological negative refraction of surface acoustic waves in a Weyl phononic crystal~\rev{\cite{he2018acoustic}}. These studies establish boundary dispersion and interface matching as control parameters for Weyl surface-wave propagation. Their extension to a high-charge Pair analogue would require a separate calculation of how the multiple surface channels couple across the interface. Reflectionlessness is not implied by IWSM irreducibility alone.

A quantum-optical extension was proposed by Garc\'ia-Elcano \textit{et al.}, who studied quantum emitters coupled to a photonic Weyl bath and used Fermi-arc propagation and negative refraction to mediate emitter interactions~\rev{\cite{garciaelcano2023quantumlink}} [\rev{Fig.~\ref{fig:pair-applications}(d)}]. Their results provide a theoretical precedent for Fermi-arc-mediated entanglement and quantum links, not an experimental demonstration of an IWSM device. A Pair-based implementation would additionally require identification of the complete mediating Weyl sector and calculation of the emitter--mode couplings. The integer incidence $F=(q)$ alone fixes neither the number of independently addressable quantum channels nor the transfer fidelity.

\subsection{Split: electronic and topological phonon candidates}
\label{sec:split_functions}

The $\{2,-1,-1\}$ Split architecture has direct antecedents in first-principles studies of topological phonons. Wang \textit{et al.} predicted a triangular Weyl complex in $\alpha$-SiO$_2$, comprising a double-Weyl point and \textcolor{red}{two same-sign unit-charge Weyl points whose combined charge compensates the double-Weyl charge} between a selected pair of phonon branches~\rev{\cite{wang2020triangular}}. The reported $P3_221$ structure (No.~154) corresponds to grey MSG 154.42 in the spinless phonon description. Ding \textit{et al.} predicted a related configuration containing a type-III charge-two Weyl phonon in BaZnO$_2$, with crystallographic space group $P3_121$ (No.~152) and corresponding grey MSG 152.34~\rev{\cite{ding2022chargetwo}}. Both MSGs are included in the single-valued Type-II entries of \rev{Table~S20 of the SM~\cite{supplemental}}. These studies therefore provide symmetry-consistent candidate topological phonon materials for the Split configuration, while experimental identification of the complete phonon Weyl inventory remains a separate task.

Electronic candidates have subsequently been proposed in light-element crystals. Zheng \textit{et al.} identified spinless $\{2,-1,-1\}$ configurations, up to global chirality reversal, in the boron allotropes ISTBN-B$_{36}$, CT-B$_{28}$, and CH-B$_{36}$~\cite{Zheng2026TNWC}. Their space groups are $P3_221$ (No.~154), $P4_1$ (No.~76), and $P6_4$ (No.~172), respectively, corresponding to grey MSGs 154.42, 76.8, and 172.126. Each belongs to the single-valued Type-II sector of \rev{Table~S20 of the SM~\cite{supplemental}}. These are electronic Weyl-semimetal predictions, not phonon realizations. The exact configurations are defined in the no-SOC limit; the small avoided crossings reported after SOC is included should not be assigned the quantized Weyl charges of the parent spinless nodes~\cite{Zheng2026TNWC}.

Bai \textit{et al.} further predicted a collinear $\{2,-1,-1\}$ electronic configuration in DZQH-C$_{36}$~\cite{Bai2026Triplet}. Its left- and right-handed structures belong to $P6_5$ (No.~170) and $P6_1$ (No.~169), respectively, with grey MSGs 170.118 and 169.114. Both settings occur in the single-valued Type-II sector of \rev{Table~S20 of the SM~\cite{supplemental}}. The reported double-Weyl node and its two compensating unit-charge nodes therefore provide another spinless electronic candidate consistent with the classified Split symmetry settings.

The heterovalent $\{3,-2,-1\}$ configuration also has a prior theoretical realization. Bai \textit{et al.} classified this triplet and constructed a twelve-band spinless tight-binding model in Type-I MSG 169.113~\cite{Bai2026Triplet}. This setting is included in \rev{Table~S23 of the SM~\cite{supplemental}}; the reported five Type-I and five Type-III supporting MSGs also agree with the corresponding entries of that table in both spin conventions. The heterovalent Split configuration should therefore not be described as previously unreported. Its evidence level nevertheless differs from the DZQH-C$_{36}$ result: the former is a symmetry classification and lattice-model realization, whereas the latter is a first-principles material prediction for $\{2,-1,-1\}$.

An experimental analogue of multibranch Weyl surface connectivity was demonstrated in an acoustic crystal by He \textit{et al.}, who observed quadratic Weyl points and double-helicoid surface arcs~\rev{\cite{he2020quadratic}}. This provides a spectral precedent for resolving the multiple surface branches associated with a higher-charge node. It should be distinguished both from an atomic-lattice phonon measurement and from an independent verification of the complete IWSM inventory.

The Split classification also motivates a specific channel-partitioning objective. For a resolved $\{3,-2,-1\}$ configuration, \rev{Sec.~\ref{sec:charge_flow}} fixes the net incidence vector to $F=(2,1)$; for $\{3,-1,-1,-1\}$, it fixes $F=(1,1,1)$. These vectors specify how many net chiral channels are associated with each compensating projection. Turning this momentum-space information into a real-space device requires additional mode-selective coupling and a defined scattering geometry.

\begin{figure}[!t]
	\centering
	\includegraphics[width=\columnwidth]{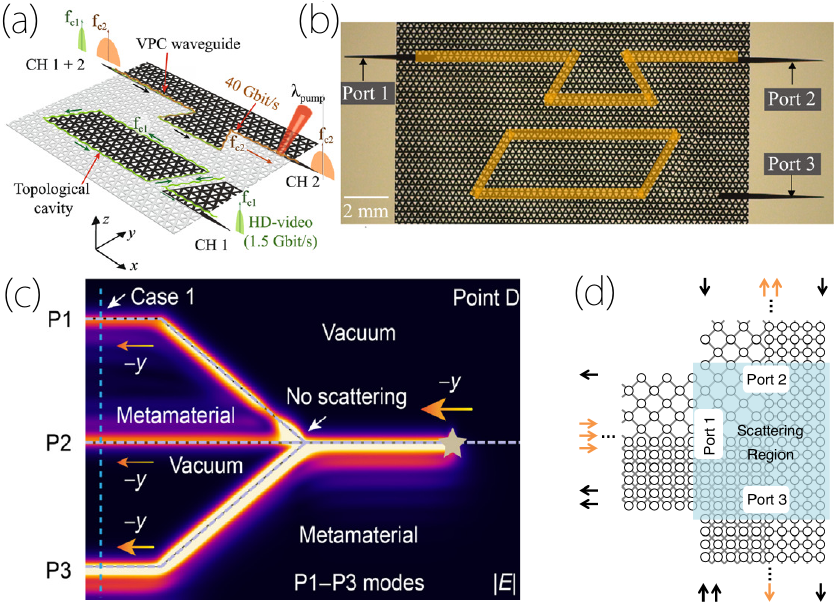}
	\caption{
		Published precedents for Split-IWSM channel control.
		(a,b) Operating schematic and fabricated topological terahertz demultiplexer; scale bar, $2\,\mathrm{mm}$.
		(c) Calculated electric-field amplitude $|E|$ for one-to-three Fermi-arc routing into outputs $P_1$--$P_3$ in a chiral Weyl metamaterial.
		(d) Theoretical Chern-photonic splitter with \rev{incoherent three-channel input} distributed into two and one outgoing channels.
		Panels (a,b), (c), and (d) are adapted from Refs.~\rev{\cite{kumar2022demultiplexer,han2025multichannel,kim2024longrange}}, respectively.
		These are experimental port-addressing and theoretical channel-routing precedents, not verified Split-IWSM implementations.
		Integer incidence does not by itself determine output-power ratios.
	}
	\label{fig:split-applications}
\end{figure}

Several published studies provide distinct parts of such an implementation. Kumar \textit{et al.} experimentally demonstrated a chip-scale topological terahertz demultiplexer based on valley-photonic waveguides and cavities~\rev{\cite{kumar2022demultiplexer}}. \rev{Its operating scheme [Fig.~\ref{fig:split-applications}(a)] and fabricated device [Fig.~\ref{fig:split-applications}(b)] show the separation of input frequencies into distinct output ports.} This establishes frequency-selective delivery to separate outputs, but not a Weyl-charge-controlled splitting rule. Han \textit{et al.} studied multichannel directional Fermi-arc routing in chiral metamaterials using numerical field distributions~\rev{\cite{han2025multichannel}} [\rev{Fig.~\ref{fig:split-applications}(c)}]. Their work supplies a Weyl-surface-wave routing precedent rather than an experimental certification of a Split IWSM.

A closer channel-counting comparison is the theoretical Chern-photonic splitter of Kim \textit{et al.}, in which \rev{incoherently averaged three-channel incidence} connects to two outgoing channels in one arm and one in another~\rev{\cite{kim2024longrange}} [\rev{Fig.~\ref{fig:split-applications}(d)}]. The reported transmission properties were obtained from a specified scattering model, including the treatment of the input channels. This distinction matters here: the incidence matrix $F$ counts net topological connections, whereas a device scattering matrix contains amplitudes and phases. Consequently, a Split-IWSM incidence vector does not by itself imply a quantized output-power ratio or conductance.

An IWSM-specific test would combine independent bulk identification with energy- or frequency-resolved tracking of the surface channels. For a port-based implementation, the correspondence between projected Weyl sectors and external channels would first have to be established in the device model. Only then could a measured scattering response be compared with the integer incidence predicted for the selected Split configuration. The cited spectroscopy, demultiplexing, and scattering studies provide precedents for these separate steps~\rev{\cite{kumar2022demultiplexer,han2025multichannel,he2020quadratic,kim2024longrange}}; their combination in one independently verified Split IWSM remains a proposed implementation rather than a result established by the present lattice models.

\subsection{\texorpdfstring{\rev{Mixed: symmetry targets and boundary matching}}{Mixed: symmetry targets and boundary matching}}
\label{sec:mixed_boundary_matching}

\auditrev{
Mixed IWSMs admit different labeled incidence matrices for the same bulk Weyl inventory and projected terminal charges. For \mbox{$\{3,1,-2,-2\}$}, the two matrices in Eq.~(\ref{eq:mixed_example_1}) connect the unit-charge source to different $C=-2$ sinks. These matchings are equivalent for unlabeled sinks but distinguishable when their surface momenta are resolved. For \mbox{$\{3,3,-2,-2,-2\}$}, six units of flow admit both all-to-all matching and the selective row patterns $(2,1,0)$ and $(0,1,2)$ [Eq.~(\ref{eq:mixed_example_2})]. This freedom motivates reconfigurable multichannel routing through changes in $F$ at fixed row and column sums.

Surface chemistry and interface geometry provide complementary ways to explore this freedom. Yang \textit{et al.} observed a surface topological Lifshitz transition in potassium-decorated NbAs~\cite{yang2019decoration}; Morali \textit{et al.} found termination-dependent arc connectivity in Co$_3$Sn$_2$S$_2$~\cite{morali2019}. Wang \textit{et al.} observed Fermi-arc reconstruction and redistribution of interface fields upon twisting two photonic Weyl meta-crystals~\cite{wang2024twisted}. These studies establish boundary-control mechanisms in related Weyl systems; NbAs, with unit-charge nodes, does not realize either Mixed inventory. For Mixed models, surface-potential tuning could test matching changes without modifying the bulk Hamiltonian. Direct comparison using momentum-labeled terminals requires conserved surface momentum; aperiodic interfaces instead require an appropriate interface-mode and scattering description. Momentum-resolved spectra and spatial field maps can then connect changes in terminal matching to their real-space propagation consequences.

Spectroscopy and transport offer complementary readouts of the resulting boundary reconstruction. Jia \textit{et al.} theoretically related Fermi-arc Lifshitz transitions to sign reversals of the longitudinal second-order nonreciprocal conductivity in Co$_3$Sn$_2$S$_2$~\cite{jia2026connectivity}. For a Mixed implementation, correlating such a response with energy-resolved surface matching would test the electrical consequences of boundary control. Wave-based implementations could instead couple distinguishable surface-channel groups to separately addressable outputs and track how a boundary perturbation redistributes an injected mode. In both cases, $F$ specifies net chiral connections; actual transmission and output powers require a boundary-specific scattering calculation.

Our lattice realizations [Sec.~\ref{sec:constructive_realizations}, Fig.~\ref{fig:mixed}] and reference surface spectra [SM~\cite{supplemental}, Fig.~S3] provide a starting point for these tests. The configuration-specific MSG and irrep/corepresentation channels in SM Tables~S24 and S25 guide the search for candidate materials. Candidate identification must use the appropriate spin convention and verify the complete Mixed inventory in the selected adjacent-band gap. Access to distinct admissible matrices must be established along a boundary-parameter trajectory while confirming unchanged bulk nodes and projected terminal charges. These tests would connect symmetry-selected Mixed inventories to reconfigurable routing with correlated spectroscopic, propagation, or electrical readout.
}

\begin{figure*}[!t]
	\centering
	\includegraphics[width=0.84\textwidth]{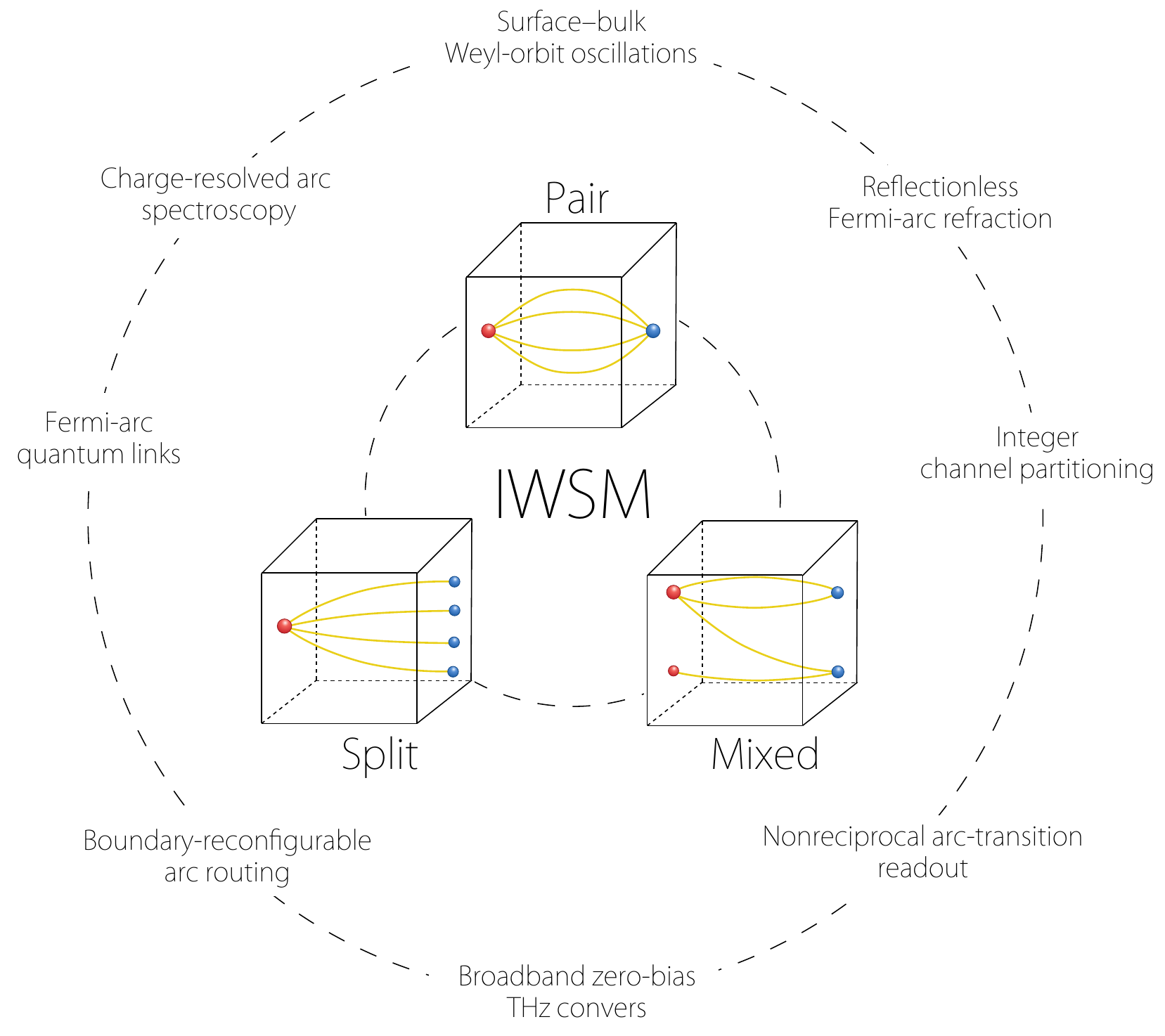}
	\caption{
		\rev{Research directions organized by Pair, Split, and Mixed IWSM architectures.
		Red and blue spheres denote opposite projected chiral charges; gold curves represent units of reduced net chiral incidence.
		The surrounding labels summarize spectroscopic, transport, refractive, quantum-link, and channel-control routes supported by the precedents in Figs.~\ref{fig:pair-applications} and \ref{fig:split-applications} and discussed in \auditrev{Secs.~\ref{sec:pair_functions}--\ref{sec:cross_motif_functions}}.
		\auditrev{Broadband zero-bias rectification is a cross-motif opportunity governed by the material's nonlinear-response symmetry~\cite{kumar2021rf,hu2026rfthz}.}
		The diagram summarizes prospective applications of verified IWSM sectors, rather than demonstrated IWSM devices or guaranteed performance.}
	}
	\label{fig:landscape}
\end{figure*}

\subsection{Cross-motif responses and symmetry-guided inverse design}
\label{sec:cross_motif_functions}

Some functionalities are controlled primarily by response symmetry rather than by the Pair, Split, or Mixed architecture. The Berry-curvature-dipole mechanism, for example, permits a second-order Hall response in appropriate inversion-breaking, time-reversal-invariant metals~\rev{\cite{sodemann2015nonlinear}}. Experiments on TaIrTe$_4$ demonstrated room-temperature nonlinear Hall response and wireless radio-frequency rectification~\rev{\cite{kumar2021rf}}; \auditrev{subsequent work demonstrated zero-bias rectification from 19 MHz to 2.88 THz and passive frequency mixing over 0.1--40 GHz~\cite{hu2026rfthz}. These measurements motivate broadband detection and rectification across radio-frequency and terahertz input bands, with mixing as a distinct functionality.} They do not demonstrate that irreducibility is necessary for the effect or that a smaller Weyl inventory automatically increases its magnitude.

An additional experimental precedent is the chiral fermionic valve reported by Dixit \textit{et al.} in PdGa, where a three-arm device spatially separates currents associated with opposite chiralities and permits their quantum interference~\rev{\cite{dixit2026valve}}. PdGa hosts multifold fermions and therefore lies outside the stable twofold-Weyl domain classified here. Its relevance is the demonstrated possibility of chirality-selective current manipulation, not an identification of a Pair, Split, or Mixed IWSM or a measurement of the incidence matrix $F$.

The configuration-resolved symmetry tables provide a more direct starting point for materials design than a search for individual high-charge WPs. \rev{Tables~S16--S19 of the SM~\cite{supplemental}} specify the allowed settings for the four Pair classes, \rev{Tables~S20--S23} for the four Split classes, and \rev{Tables~S24 and S25} for the two Mixed classes. For each configuration, the catalogue already identifies candidate MSGs, the applicable single- or double-valued representation convention, the allowed crossing locations, and the associated irrep/corepresentation channels. The inverse problem can therefore begin with a desired complete signed Weyl configuration and its boundary incidence, rather than with an unconstrained material pool.

The first design choice is the target IWSM class and its physical representation convention. For a spinless electronic approximation or a topological phonon realization, the relevant candidates are drawn from the single-valued entries. For a spinful electronic phase with SOC, the search must instead use the double-valued entries and a magnetic order compatible with the specified MSG. These two searches cannot be interchanged by simply retaining the same crystallographic space-group number. The spinless and finite-SOC results for the nonmagnetic boron candidates illustrate why the representation convention is part of the physical design constraint~\rev{\cite{zheng2026boron,Zheng2026TNWC}}.

The next step is to use the tabulated crossing channels to select a symmetry-compatible band-connectivity pattern. For example, the $I4_1$ carbon candidate connects the $\{2,-2\}$ material search to the single-valued grey-group entry 80.30 in \rev{Table~S17 of the SM~\cite{supplemental}}, while the electronic Split candidates discussed above map to specific entries of \rev{Table~S20}. For the heterovalent $\{3,-2,-1\}$ class, \rev{Table~S23} supplies five Type-I and five Type-III settings in each spin convention, including 169.113 and 178.159. Its sixfold-screw crossing channels identify the required coexistence of charge-one, charge-two, and charge-three nodes on the same symmetry axis. The design target is therefore not merely a sixfold crystal or a triple-Weyl point, but the complete symmetry-compatible three-node configuration.

Given such a target, crystal-structure and electronic-structure searches can be restricted to candidates with the appropriate MSG and active-band representations. First-principles calculations must then determine whether the required band ordering and crossings actually occur near the Fermi level, whether SOC and the selected magnetic order preserve them, and whether additional crossings occur between the same adjacent bands. For candidate topological phonon materials, the corresponding checks concern structural stability, the relevant phonon branches, and their complete Weyl-point configuration. The tabulated symmetry conditions restrict the search space but do not determine the chemical energetics or guarantee that every material in an allowed MSG realizes the prescribed phase.

The final validation combines the complete BZ node search with enclosing-surface Chern-charge calculations and the projected surface spectral flow. Pair candidates require identification of one complete opposite-charge pair. Split candidates require the terminal-resolved integer incidences of the chosen configuration. Mixed candidates require both bulk certification and reconstruction of the surface matching before a boundary-induced change can be assigned to the classified architecture. Charge-resolved measurements in related photonic and acoustic systems provide experimental precedents for the spectral part of this procedure~\rev{\cite{chen2022chargefour,he2020quadratic}}, while any device implementation additionally requires a material- and boundary-specific coupling model.

\rev{Symmetry-guided discovery and functional exploration provide complementary routes toward verified irreducible Weyl sectors [Fig.~\ref{fig:landscape}].} The cited literature provides electronic and topological phonon candidates for several configurations, as well as a prior lattice-model realization of the heterovalent Split class. Other configurations remain candidates for material discovery even when their symmetry conditions and model realizations are known. \rev{Tables~S16--S25 of the SM~\cite{supplemental}} turn these searches into configuration-specific problems by supplying the candidate magnetic symmetries and crossing representations in advance. The published spectroscopic, transport, refractive, and channel-control mechanisms then define possible measurements and applications of a verified irreducible Weyl sector. Their quantitative performance remains a separate question that cannot be inferred from the signed Weyl inventory alone.

\section{Discussion}
\label{sec:discussion}

An IWSM is distinguished by the completeness of its neutral Weyl configuration, not by a new local charge assigned to an individual node. Local symmetry analysis determines the possible Weyl crossings, and band-representation theory constrains their connectivity~\rev{\cite{fang2012multiweyl,watanabe2018msg,elcoro2021mtqc,bradlyn2017}}. The present criterion additionally requires that one primitive IWM exhaust the crossings of a selected adjacent-band pair near the Fermi level, with exact MSG-orbit closure and compatible band representations.

This requirement separates the ten IWSM classes from the sixteen-IWM generating framework~\rev{\cite{pang2026universal}}. The sixteen modules describe charge-neutral Weyl assemblies; the ten classes identify those modules that admit standalone crystalline realizations under the stronger selected-band criterion. The remaining modules are not excluded from the charge decomposition of larger configurations. Nor does an algebraic IWM decomposition imply that a Bloch Hamiltonian can be decoupled into independent IWSM blocks.

Earlier single-pair and triplet constructions are included within this framework~\rev{\cite{wang2019eucd2as2,wang2020triangular,wang2022singlepair,Bai2026Triplet}}. In particular, $\{3,-2,-1\}$ already has a symmetry classification and a lattice-model realization~\cite{Bai2026Triplet}. Its comparison with $\{3,-3\}$ and $\{3,-1,-1,-1\}$ shows why the compensating nodes matter: the same local triple-Weyl charge enters different complete configurations and different boundary incidences. Irreducibility is retained under symmetry-compatible deformations that preserve the signed inventory and its complete MSG orbits while leaving the selected bands nondegenerate away from the member nodes.

For resolved projections and complete reduced incidence, primitivity requires a connected boundary graph. Pair and Split configurations fix scalar and vector incidences, whereas Mixed configurations allow several labeled matrices with the same margins. These constraints do not fix microscopic Fermi-arc contours. Boundary modification can reconnect surface states without changing bulk nodes~\rev{\cite{devizorova2017,morali2019,yang2019decoration,dwivedi2016tunable}}, and a reducible bulk configuration can consequently exhibit connected surface incidence. Projection overlap or unresolved spectral weight can obstruct the boundary diagnostic without changing bulk irreducibility.

Identification therefore requires the complete selected-band Weyl inventory, its signed charges, and its actual MSG assignment, followed by an appropriate boundary test. Energy- or frequency-resolved spectroscopy can examine net chiral spectral flow, as illustrated by measurements of quadratic and charge-four Weyl systems~\rev{\cite{chen2022chargefour,he2020quadratic}}. A visible arc or connected constant-energy contour is insufficient. Likewise, the integer matrix $F$ is not a device scattering matrix; relating it to conductance or output power requires a separate coupling and scattering calculation.

A proposed reducible control for $\{3,-2,-1\}$ is the charge inventory
\begin{equation}
	\mathcal{C}_{\rm ref}
	=
	\{3,-3\}
	\uplus
	\{2,-2\}
	\uplus
	\{1,-1\},
	\label{eq:discussion_reducible_reference}
\end{equation}
where $\uplus$ retains all members of the charge multisets. The reference preserves the selected local charge types but adds partners that permit independent neutral sectors. A decoupled implementation would have block-diagonal boundary incidence, unlike the primitive Split configuration. Realizing a quantitative control with comparable dispersions and boundary conditions remains a separate task, rather than a result established by this charge-level construction.

\rev{Tables~S16--S25 of the SM~\cite{supplemental}} specify symmetry-resolved starting points for such model comparisons and for materials searches. Their single- and double-valued settings constrain candidate symmetries and crossing representations, but do not guarantee the required band ordering in a particular compound. The electronic definition concerns the selected bands near the Fermi level; other pockets or other-band crossings can affect the full material response without changing that sector. Topological phonon and photonic analogues instead use the corresponding Hermitian branch structure.

The ten classes exhaust the standalone primitive inventories within the stated stable twofold-Weyl domain $|C|\leq4$~\cite{Yu2022TypeII,Liu2022TypeIII,Zhang2022TypeIV,zhang2020quadruple,cui2021chargefour}, not all Hamiltonians or all adiabatic equivalence classes. Multifold crossings, extended degeneracies in the selected pair, interacting zeros, and non-Hermitian exceptional structures require separate classifications. Within this scope, the framework distinguishes the irreducibility of a complete crystalline Weyl configuration from both local-node topology and boundary-dependent spectral details.

\section{Conclusion}
\label{sec:conclusion}

We have classified irreducible Weyl semimetals (IWSMs) by requiring the complete Weyl-node configuration of a selected adjacent-band pair near the Fermi level to realize one primitive irreducible Weyl molecule (IWM). Combining charge primitivity, exact MSG-orbit closure, and band-representation compatibility across all 1651 MSGs identifies ten standalone IWSM classes among the sixteen IWM generators, within the stable twofold-Weyl domain $|C|\leq4$. Remarkably, five of these classes are previously unrecognized: $\{3,-3\}$, $\{3,-1,-1,-1\}$, $\{4,-1,-1,-1,-1\}$, $\{3,1,-2,-2\}$, and $\{3,3,-2,-2,-2\}$. Explicit single-valued lattice models realize all ten classes. The four Pair, four Split, and two Mixed architectures distinguish complete Weyl configurations that cannot be characterized by their local node charges alone. Under resolved projection and the reduced net-flow assumptions, irreducibility requires connected boundary incidence. Pair and Split fix scalar and vector incidences, whereas Mixed permits nonunique labeled matrices with fixed terminal charges; microscopic Fermi-arc contours remain boundary dependent. The single- and double-valued symmetry catalogues provide configuration-specific targets for materials discovery, while identification requires the complete bulk Weyl inventory and its boundary spectral-flow test. This framework connects primitive charge-neutral building blocks to independent crystalline Weyl phases and their topological phonon and photonic analogues.

\section*{Data and Code Availability}

All classification records and model evidence supporting this study are provided in the article and the Supplemental Material. These include symmetry and compatibility catalogues, model Hamiltonians and parameters, Weyl-point coordinates, \WCC{} diagnoses, and surface spectra. No separate data or code package accompanies this submission. The external symmetry-analysis and numerical packages used in this work are available through the cited references~\cite{Shinohara2023Spglib,Liu2023MSGCorep,Zhang2022MagneticTB,gresch2017z2pack,wu2018wanniertools}.

\section*{Figure Credits}

\rev{The source panels adapted in Figs.~\ref{fig:pair-applications} and \ref{fig:split-applications}, as identified in their captions, are licensed under the \href{https://creativecommons.org/licenses/by/4.0/}{Creative Commons Attribution 4.0 International License}. The source panels were cropped, rescaled, and rearranged and, where applicable, relabeled; no plotted data were altered.}

\begin{acknowledgments}
The authors thank Weikang Wu and Shengyuan A. Yang for valuable discussions. This work was supported by the National Natural Science Foundation of China (Grants No.~12304202), the Hebei Natural Science Foundation (Grant No.~A2023203007), and the Science Research Project of Hebei Education Department (Grant No.~BJK2024085).
\end{acknowledgments}

\nocite{Liu2022TypeIII,Tang2021EffectiveModels,Zhang2022TypeIV,Zhao2016Nonsymmorphic,hirschmann2021tetragonal}

\end{document}